\documentclass[twocolumns]{emulateapj} 
 
\usepackage{verbatim} 
\usepackage{natbib} 
\usepackage{amsmath} 
\usepackage{amsbsy} 
\usepackage{lscape} 
\usepackage{epstopdf} 
 
\slugcomment{Draft} 
\shorttitle{Lyman-$\alpha$ from Cold Accretion} 
\shortauthors{Faucher-Gigu\`ere et al.} 
 
\newcommand{\Lya}{\mbox{Ly$\alpha$}}

\begin{document} 

\title{LY$\alpha$ Cooling Emission from Galaxy Formation} 
 
\author{Claude-Andr\'e Faucher-Gigu\`ere\altaffilmark{1,2,$\star$}, Du\v{s}an Kere\v{s}\altaffilmark{1,2,$\dagger$}, Mark Dijkstra\altaffilmark{1,3}, Lars Hernquist\altaffilmark{1}, Matias Zaldarriaga\altaffilmark{4}} 
\altaffiltext{1}{Department of Astronomy, Harvard University, Cambridge, MA 02138, USA.} 
\altaffiltext{2}{Department of Astronomy and Theoretical Astrophysics Center, University of California, Berkeley, CA 94720-3411, USA.} 
\altaffiltext{3}{Max-Planck-Institut f\"ur Astrophysik, Karl-Schwarzschild-Strasse 1, 85748 Garching, Germany.} 
\altaffiltext{4}{School of Natural Sciences, Institute for Advanced Study, Princeton, NJ 08540, USA.}
\altaffiltext{$\star$}{Miller Fellow; cgiguere@berkeley.edu} 
\altaffiltext{$\dagger$}{Hubble Fellow}
 
\begin{abstract}
Recent numerical and analytical studies have shown that galaxies accrete most of their baryons via the cold mode, from streams with temperatures $T\sim10^{4}-10^{5}$ K. 
At these temperatures, the streams should radiate primarily in the \Lya~line and have therefore been proposed as a model to power the extended, high-redshift objects known as \Lya~blobs, and may also be relevant for powering a range of less luminous \Lya~sources. 
We introduce a new \Lya~radiative transfer code, $\alpha RT$, and calculate the transport of the \Lya~emission from cold accretion in cosmological hydrodynamical simulations.  
In this paper, we describe our methodology, and address physical and numerical issues that are critical to making accurate predictions for the cooling luminosity, but that have been mostly neglected or treated simplistically so far. 
In particular, we highlight the importance of self-shielding and of properly treating sub-resolution models in numerical simulations.   
Most existing simulations do not self-consistently incorporate these effects, which can lead to order-of-magnitude errors in the predicted cooling luminosity. 
Using a combination of post-processing ionizing radiative transfer and re-simulation techniques, we develop an approximation to the consistent evolution of the self-shielded gas. 
We quantify the dependence of the \Lya~cooling luminosity on halo mass at $z=3$ for the simplified problem of pure gas accretion embedded in the cosmic radiation background and without feedback, and present radiative transfer results for a particular system. 
While pure cooling in massive halos (without additional energy input from star formation and AGN) is in principle sufficient to produce $L_{\alpha}\sim10^{43}-10^{44}$ erg s$^{-1}$ blobs, this requires including energy released in gas of density sufficient to form stars, but which is kept 100\% gaseous in our optimistic estimates. 
Excluding emission from such dense gas yields lower luminosities by up to one to two orders of magnitude at high masses, making it difficult to explain the observed \Lya~blobs with pure cooling.  
Resonant scattering produces diffuse \Lya~halos, even for centrally concentrated emission, and broad double peaked line profiles. 
In particular, the emergent line widths are in general not representative of the velocity dispersion within galactic halos and cannot be directly used to infer host halo masses. 
\end{abstract}

\section{INTRODUCTION}
\label{introduction}
\cite{2000ApJ...532..170S} discovered two extremely bright, large, and diffuse \Lya-emitting `blobs' in a narrowband survey of the SSA 22 proto-cluster region at $\langle z \rangle=3.09$ \citep[for the early detection of extended \Lya~emission at $z\sim2.4$, see also][]{1996ApJ...457..490F, 2001ApJ...554.1001F, 1999AJ....118.2547K}.
These two blobs, labelled LAB1 and LAB2, have physical extent $\gtrsim140$ kpc, are more luminous ($L_{\rm Ly\alpha}\sim10^{44}$ erg s$^{-1}$) than typical line emitters at the same redshift by a factor $\sim10-100$, and unlike similar halos around radio galaxies have no detectable radio continuum.
Since their discovery, these blobs have become some of the most spectacular of a new class of sources that now counts several tens of members \citep[e.g.,][]{2004AJ....128..569M, 2004ApJ...602..545P, 2004ApJ...614...75F, 2005ApJ...629..654D, 2006A&A...452L..23N, 2006ApJ...648...54S, 2007MNRAS.378L..49S, 2008ApJ...678L..77P, 2009ApJ...702..554P, 2009ApJ...696.1164O, 2009ApJ...693.1579Y, 2010ApJ...719.1654Y} and whose nature remains unclear.
Because detecting spatially extended \Lya~emission usually requires deep narrowband imaging, which covers thin redshift slices, only a small volume the Universe has been effectively surveyed for them to date; the blobs and their fainter analogues are therefore likely to be numerous and cosmologically significant.
In fact, existing observations and theoretical models indicate that they may be the sites of massive galaxy formation and also may display signatures of associated active galactic nuclei (AGN) and/or supernova feedback.
The \Lya~blobs thus provide a unique opportunity to probe the processes driving and regulating galaxy formation, and may be intimately related to phenomena including proto-clusters, mergers, and submillimeter galaxies.\\ \\
Although the most extreme \Lya~blobs have received the most attention, there in fact exists a wide continuum of spatially extended \Lya~sources at high redshift, for example the ones discovered by \cite{2006ApJ...648...54S} with line luminosities $\sim10^{42}$ erg s$^{-1}$ and those discovered by \cite{2008ApJ...681..856R}, with line luminosities as low as $\sim10^{39}$ erg s$^{-1}$. 
Understanding the nature of these fainter but more numerous sources is equally important to develop a physical picture of galaxy formation. 
In fact, the fainter sources likely probe different (perhaps earlier) stages of galaxy assembly.\\ \\
A central puzzle for the \Lya~blobs is that many of them do not appear to have a central source energetic enough to power their entire \Lya~emission \citep[e.g.,][]{2004AJ....128..569M, 2006A&A...452L..23N}.
Even for the blobs which do have energetically-sufficient counterparts \citep[e.g.,][]{2005MNRAS.363.1398G, 2007ApJ...655L...9G, 2009ApJ...700....1G, 2009ApJ...692.1561W}, it is unclear whether that energy can actually couple effectively to the \Lya~emission.
Submillimeter starbursts and obscured AGN imply the presence of large quantities of dust, which acts to destroy \Lya~photons particularly efficiently \citep[][]{1990ApJ...350..216N}.
It is therefore uncertain whether \Lya~photons produced by such dust-enshrouded sources can escape in significant amounts.  
Nevertheless, several different mechanisms been proposed to power the blobs and can be broadly divided into three categories:\\ \\
{\bf Embedded star formation or AGN}, possibly obscured from direct view by dust, could photoionize the surrounding hydrogen nebula \citep[][]{1998MNRAS.299..661M, 2001ApJ...556...87H, 2004Natur.430..999W, 2005A&A...436..825W, 2007ApJ...657L..69L}.\\ \\
{\bf Superwinds} driven by starburst supernovae could explain the observed sizes and kinematics of the blobs, with the \Lya~emission being generated in the swept up material \citep[][]{2000ApJ...532L..13T, 2001ApJ...562L..15T, 2003ApJ...591L...9O, 2004ApJ...613L..97M, 2005Natur.436..227W, 2005MNRAS.363.1398G}.
AGNs could also drive similar winds \citep[e.g.,][]{2005ApJ...618..569M, 2005MNRAS.361..776S, 2005Natur.433..604D, 2006ApJS..166....1H}.\\ \\
{\bf Cooling radiation}, emitted as gas accretes onto forming galaxies, could produce luminous and extended structures.
If a large fraction of the accreting gas has a temperature $T\sim10^{4}-10^{5}$ K, then most of the cooling radiation could be \Lya~\citep[][]{1977MNRAS.180..479F, 1991ApJ...377..365K, 1992ApJ...391..608H, 2000ApJ...537L...5H, 2001ApJ...562..605F, 2003MNRAS.345..349B, 2005MNRAS.363....2K, 2005ApJ...622....7F, 2006ApJ...649...14D, 2006ApJ...640..539Y, 2009MNRAS.400.1109D, 2010MNRAS.407..613G, 2010MNRAS.402.1449D}.\\ \\
In this work, motivated by recent progress on our understanding of how galaxies get their gas, we focus on the third possibility. 
The methods developed could however be applied to the first two classes of models as well, and we plan to extend our calculations to model those processes in the future.\\ \\
In the classic sketch of galaxy formation \citep[][]{1977MNRAS.179..541R, 1977ApJ...211..638S, 1978MNRAS.183..341W}, gas falling into dark matter halos is shocked and heated to the virial temperature.
For a galaxy with a mass similar to that of the Milky Way, the shocked gas attains a temperature $T_{\rm vir}\sim10^{6}$ K.
In the dense inner regions of the halos, this gas efficiently radiates its thermal energy, loses its pressure support, and settles into compact discs where it can form stars.
A wealth of recent work however suggests that this picture requires an important modification: most of the gas is never strongly shocked as it flows toward the central forming galaxy, but rather accretes in a ``cold mode'', maintaining a temperature $T\lesssim10^{5}$ K.
Moreover, this cold accretion proceeds through dense filaments rather than in a spherically symmetric fashion.\\ \\
Although the importance of cold accretion in galaxy formation has only recently been demonstrated in high-resolution three-dimensional hydrodynamical simulations \citep[e.g.,][]{2001ApJ...562..605F, 2003ASSL..281..185K, 2005MNRAS.363....2K, 2009MNRAS.395..160K, 2008MNRAS.390.1326O, 2009ApJ...694..396B, 2009Natur.457..451D}, \cite{1977ApJ...215..483B} had argued on the basis of analytic models of protogalaxy collapse that the amount of shock heating could be small for plausible physical conditions.
Moreover, already in the first simulations of forming galaxies, most of the gas never heated above $T\sim3\times10^{4}$ K \citep[][]{1991ApJ...377..365K} and the importance of filamentary structures was recognized by \cite{1993ApJ...412..455K} and \cite{1994MNRAS.270L..71K}.
\cite{2003MNRAS.345..349B} carried out a stability analysis, supported by one-dimensional hydrodynamical simulations \citep[see also][]{2007MNRAS.380..339B, 2006MNRAS.368....2D}, and found that when the radiative cooling is efficient compared with the infall rate, the post-shock gas becomes unstable and cannot support the shock.
When applied to cosmology, their results agree well with those of three-dimensional simulations.\\ \\
The \Lya~emission from cold accretion has already been the subject of some studies. 
\cite{2001ApJ...562..605F} first evaluated the \Lya~cooling luminosity from hydrodynamical simulations and suggested that it could account for the \Lya~blobs discovered by \cite{2000ApJ...532..170S}. 
\cite{2000ApJ...537L...5H} reached a similar conclusion using simplified analytic arguments. 
Also using simulations, \cite{2005ApJ...622....7F} found that the \Lya~cooling radiation from structure formation could account for some, but not all, of the luminosity of the \Lya~blobs. 
\cite{2006ApJ...640..539Y} studied both the hydrogen and helium cooling radiation using simulations, but found the hydrogen \Lya~luminosity to be strongly dependent on the self-shielding correction applied. 
Recently, \cite{2009MNRAS.400.1109D} developed an analytic model and suggested that cooling radiation from the cold mode could account for all the \Lya~blobs under reasonable assumptions.  
Using adaptive mesh refinement (AMR) simulations, \cite{2010MNRAS.407..613G} provided supporting evidence for this picture. 
In our discussion (\S \ref{discussion}), we will contrast our main results with those of \cite{2010MNRAS.407..613G}, concluding that the differences with theirs most likely originate from the treatments of self-shielding and sub-resolution modeling, which are a focus of our study.\\ \\
No study focusing specifically on cooling emission has however combined realistic \Lya~radiative transfer with hydrodynamical simulations before.\footnote{Other authors have included a cooling component in radiative transfer calculations of \Lya-emitting galaxies \citep[e.g.,][]{2006ApJ...645..792T, 2007ApJ...657L..69L, 2009ApJ...696..853L, 2009ApJ...696..853L}, but have not explicitly separated out the signatures of pure cooling or investigated the important uncertainties in detail.}Because \Lya~photons resonantly scatter, the resultant morphologies, spatial extents, and spectra are strongly modified by radiative transfer effects \citep[e.g.,][]{2006ApJ...649...14D}. 
Since the \Lya~photons tend to follow paths of least resistance in space and frequency (\S \ref{lya radiative transfer}), the spatial geometry and bulk velocity fields play critical roles in determining the radiation transport \cite[for observational evidence of these effects, see e.g.][]{1998A&A...334...11K, 2003ApJ...598..858M}. 
Fully three-dimensional calculations are therefore necessary to make realistic predictions. 
Moreover, both the existing analytical and numerical studies have limitations that could induce important errors in quantities as basic as the integrated \Lya~luminosity of the cold streams. 
One such uncertainty arises from the exponential dependence of the \Lya~emissivity on the gas temperature for the temperatures $T \sim 10^{4}$ K that are characteristic of cold accretion (\S \ref{lya emission}). 
At present, most galaxy formation simulations do not self-consistently predict the temperature distribution within the streams. 
In fact, existing simulations usually do not follow the transport of the ultra-violet (UV) radiation that ionizes and heats dense gas. 
This seriously limits the predictive power of these calculations, since small errors in the temperatures can result in large errors in the \Lya~cooling emission, and potentially grossly violate energy conservation. 
As we will demonstrate, models of sub-resolution physics in hydrodynamical simulations can also introduce large errors if not properly taken into account. 
Analytical studies based on energetic considerations are not as sensitive to the temperature of the cold streams \citep[e.g.,][]{2009MNRAS.400.1109D}, but are not immune of uncertainties either, since they rely on assumptions regarding the efficiency of \Lya~emission. 
Moreover, their simplified nature does not lend itself to detailed radiative transfer predictions.   
Resolving these issues is critical to relating the \Lya~emission from cold accretion to observations.\\ \\
Our ultimate goal is a systematic investigation of the \Lya~emission from galaxy formation that is both detailed in its predictions, and robust. 
By detailed, we envision predictions that can be directly compared with observations, and therefore require both 3D hydrodynamical simulations and realistic radiative transfer. 
By robust, we mean that the predictions should be free of assumptions that introduce the kind of large uncertainties that existing studies are subject to. 
Due to the complexity of the problem, this ultimate goal is likely to require a long-term effort.
The present paper is dedicated to laying down some of the foundations for this research program. 
Specifically, we present a new \Lya~radiative transfer code, named $\alpha RT$, and describe its application to the cooling radiation in cosmological simulations of galaxy formation \citep[for other applications of \Lya~radiative transfer codes to hydrodynamical simulations, see e.g.][]{2005ApJ...628...61C, 2006ApJ...645..792T, 2007ApJ...657L..69L, 2009ApJ...696..853L, 2010ApJ...708.1048K, 2010ApJ...716..574Z}. 
We pay particular attention to clarifying the physical and numerical uncertainties of these calculations, in particular with respect to the predicted \Lya~luminosities, and illustrate the importance of radiative transfer effects. 
To do so, aside for calculating the \Lya~luminosities of a sample of halos from a cosmological volume, we focus our radiative transfer calculations on a particular system at $z=3$ and explore variations in both the emission and radiative transfer physics. 
We limit ourselves to the most basic physical problem of accreting halos embedded in a cosmic ionizing background and neglect feedback processes. 
Follow up studies will build on the results obtained here and investigate the properties of the \Lya~emission as a function of halo mass and redshift, and will extend them by incorporating additional physics, including feedback \citep[][]{cfcosmo}.\\ \\
We begin by describing our hydrodynamical simulations in \S \ref{simulations}.
We address the emission of \Lya~photons in \S \ref{lya emission} and explicitly demonstrate the sensitivity of the predicted \Lya~cooling luminosity on assumptions regarding the thermal state of the self-shielded gas. 
Using a combination of post-processing ionizing radiative transfer and re-simulation techniques, we develop an approximation to consistently model the evolution of the dense gas in the hydrodynamical simulations, resulting in the most robust numerical predictions to date. 
In \S \ref{lya radiative transfer}, we present the results of \Lya~radiative transfer calculations for a particular system of total mass $M_{h}=2.5\times10^{11}$ M$_{\odot}$ at $z=3$ and highlight the role of both the bulk velocity flows and of the resonant scatters in shaping the emergent morphology and spectrum. 
Finally, we discuss our results and conclude in \S \ref{discussion}.
The Appendices document the radiative transfer code $\alpha RT$ introduced in this work, the ionizing radiative transfer method, and relevant analytical estimates.\\ \\
Throughout, we assume a cosmology with $(\Omega_{\rm m},~\Omega_{\rm b},~\Omega_{\Lambda},~h,~\sigma_{8},~n_{\rm s})=(0.28,~0.046,~0.72,~0.70,~0.82,~0.96)$, as inferred from the \emph{Wilkinson Microwave Anisotropy Probe} (WMAP) five-year data in combination with baryon acoustic oscillations and supernovae \citep[][]{2009ApJS..180..330K}.
While some of our hydrodynamical simulations were run with slightly different parameters, none of our conclusions are sensitive to the details of the cosmology.
We assume hydrogen and helium mass fractions of $X=0.75$ and $Y=0.25$ \citep[e.g.,][]{2001ApJ...552L...1B}, the collisional ionization coefficients given in \cite{1996ApJS..105...19K}, the \Lya~collisional excitation coefficient and average number of \Lya~photons produced per recombination from \cite{2006agna.book.....O}, and the recombination coefficients in the appendix of \cite{1997MNRAS.292...27H}.
For convenience, some symbols used in this work are defined in Table \ref{definitions}.

\begin{deluxetable*}{ll}
\tablewidth{0pc}
\tablecaption{Symbols used in this work\label{definitions}}
\tabletypesize{\footnotesize}
\tablehead{\colhead{Symbol} & \colhead{Definition}}
\startdata
$n_{i}$                   & number density of species $i$ \\
$N_{i}$                   & column density of species $i$ \\
$T$                       & gas temperature \\
$\tau_{\nu}$              & optical depth at frequency $\nu$ \\
$\nu_{0}$                 & \Lya~central frequency \\
$\Delta \nu_{\rm D}$      & \Lya~Doppler width \\
$x$                       & dimensionless frequency offset $(\nu-\nu_{0})/\Delta \nu_{\rm D}$ \\
$\Gamma_{i}$              & photoionization rate of species $i$ \\
$\Gamma_{i,c}$            & collisional ionization coefficient of species $i$ \\
$C_{\rm Ly\alpha}$         & \Lya~collisional excitation coefficient \\
$\epsilon_{\alpha}$       & \Lya~emissivity \\
$\alpha_{i}^{A,B}$\tablenotemark{a}        & case A, B recombination coefficients to species $i$ \\
$x_{\rm HI}, x_{\rm HII}$              & fractions of hydrogen in HI, HII \\
$y_{\rm HeI},~y_{\rm HeII},~y_{\rm HeIII}$ & fractions of helium in HeI, HeII, HeIII
\enddata
\end{deluxetable*} 

\section{SIMULATIONS}
\label{simulations}

\subsection{Code Details}
\label{hydro}
We compute the hydrodynamics of forming galaxies in a $\Lambda$CDM universe using a modified version of the GADGET cosmological simulation code \citep[][]{2005MNRAS.364.1105S}.
The calculation of the gravitational force uses a combination of the particle mesh algorithm \citep[e.g.,][]{1988csup.book.....H} for large separations and the hierarchical tree algorithm \citep[e.g.,][]{1986Natur.324..446B, 1987ApJS...64..715H} at small distances.
The gas dynamics is calculated using a smoothed particle hydrodynamics (SPH) algorithm \citep[e.g.,][]{1977AJ.....82.1013L, 1977MNRAS.181..375G} that conserves both energy and entropy \citep[][]{2002MNRAS.333..649S}.
The modifications with respect to the public version of the code include the treatment of cooling, the effects an uniform ultra-violet background (UVB), and a multiphase star formation algorithm as in \citet[][]{2003MNRAS.339..289S}.
Star formation is implemented by the stochastic spawning of collisionless star particles by the gas particles.
In practice, star formation in the multiphase model occurs above a density threshold of $n_{\rm H}=0.13$ cm$^{-3}$ and is calibrated to the observed \cite{1998ApJ...498..541K} law, although it plays only a tangential role in this work, which focuses  the cooling emission. 
The thermal and ionization properties of the gas are calculated including all relevant processes in a plasma with primordial abundances of hydrogen and helium following \cite{1996ApJS..105...19K}.

\subsection{Simulation Parameters and Halo Identification}
\label{simulation parameters and halo identification}
We use two types of simulations.
To achieve high resolution, we `zoom in' on individual halos within a larger simulation box and only follow the local gas dynamics at the refined resolution. 
This is done by first running a dark matter only simulation and selecting halos of interests. 
The simulation is then rerun including gas particles, with 8 times the original mass resolution, in a Lagrangian volume surrounding the halo of interest \citep[e.g.,][]{1993ApJ...412..455K}. 
In this work, we focus on zoom in simulations of an individual halo (labeled A1) selected from a volume of side length 10 $h^{-1}$ comoving Mpc. 
The A1 halo has a total mass $2.5\times10^{11}$ M$_{\odot}$ at $z=3$. 
As it is also important to understand the trends and variance between different halos, we simulated an entire cosmological volume consisting of a cubical box with a side length of 40 $h^{-1}$ comoving Mpc.
While the resolution in this volume is more limited by computational constraints, it provides us with a large number of halos of different masses.
Table \ref{simulations table} lists the simulations used in this work and their parameters. 
The minimum gas smoothing length is set to 0.1 of the gravitational softening in all the simulations.  
Given the importance of self-shielding (\S \ref{lya emission}), we rerun our simulations with exactly the same parameters, but with the UVB artificially turned off in regions exceeding a certain density (suffix $\_$ssUV), to be discussed in \S \ref{self shielding}. 
All the simulations are also rerun with the star formation model turned off. 
The simulations without star formation are identified by the additional suffix $\_$noSF. 
The A1 zoom in simulations assume a variant of the \cite{1996ApJ...461...20H} model of the UVB, while our cosmological simulations use the more recent model of \cite{2009ApJ...703.1416F}.\\ \\
A friends-of-friends (FoF) algorithm \cite[e.g.,][]{1985ApJ...292..371D} with linking length set to $b=0.2$ in units of the mean interparticle separation is used to identify the dark matter halos in the simulations.
The total mass of the particles within each FoF group, $M_{\rm FoF}$, corresponds approximately to the mass in a sphere of mean interior density 180 times the background matter density, $M_{180b}$ \citep[][]{2002ApJS..143..241W}.
We therefore define the virial radius of a halo as the radius of a sphere containing $M_{180b}$, $r_{\rm vir}\equiv r_{180b} \approx [M_{\rm FoF} / 240 \pi \rho_{u}(z)]^{1/3}$, where $\rho_{u}(z)=\rho_{\rm crit}\Omega_{m}(1+z)^{3}$ and $\rho_{\rm crit}$ is the critical density at $z=0$.
In what follows, we use $M$ as a shorthand for $M_{\rm FoF}\approx M_{180b}$.
The center of each halo is determined as the point deepest in the gravitational potential.
Since we run the FoF algorithm on the dark matter particles only, we multiply the returned masses by $\Omega_{m}/(\Omega_{m}-\Omega_{b})$ to account for the baryons.\\ \\
As we want to isolate the cold accretion cooling radiation, the simulations studied in this work do no include galactic winds or AGN feedback.
It is of course likely that the results would be somewhat modified if these processes were included.
We plan to investigate the effects of feedback and their observational manifestations in future work.

\begin{deluxetable*}{lcccccl}
\tablewidth{0pc}
\tablecaption{Hydrodynamical Simulations\label{simulations table}}
\tabletypesize{\footnotesize}
\tablehead{\colhead{Name} & \colhead{$L$ ($h^{-1}$ Mpc)\tablenotemark{a}} & \colhead{$N$\tablenotemark{b}} & \colhead{$\epsilon$ ($h^{-1}$ kpc)\tablenotemark{c}} & \colhead{Description}}
\startdata
z10n128\_A1\tablenotemark{d}                            & 10 & $2\times256^{3}$ & 0.8 & 8$\times$ mass refinement zoom \\
z10n128\_A1\_noSF               & 10 & $2\times256^{3}$ & 0.8 & no star formation \\
z10n128\_A1\_ssUV                & 10 & $2\times256^{3}$ & 0.8 & UVB off at $n_{\rm H}>0.01$ cm$^{-3}$ \\
z10n128\_A1\_ssUV\_noSF   & 10 & $2\times256^{3}$ & 0.8 & UVB off at $n_{\rm H}>0.01$ cm$^{-3}$, no SF \\
\hline
gdm40n512\tablenotemark{e}      & 40 & $2\times512^{3}$ & 1.6 & full box \\ 
gdm40n512\_noSF      & 40 & $2\times512^{3}$ & 1.6 & no star formation \\
gdm40n512\_ssUV      & 40 & $2\times512^{3}$ & 1.6 & UVB off at $n_{\rm H}>0.01$ cm$^{-3}$ \\
gdm40n512\_ssUV\_noSF      & 40 & $2\times512^{3}$ & 1.6 & UVB off at $n_{\rm H}>0.01$ cm$^{-3}$, no SF
\enddata
\tablenotetext{a}{Comoving box side length.}
\tablenotetext{b}{Effective total number of dark matter+gas particles in the box, after zoom refinement.}
\tablenotetext{c}{Comoving Plummer equivalent gravitational softening length, after zoom refinement.}
\tablenotetext{d}{Dark matter, gas, and stellar particle masses are $4\times10^{6}$ $h^{-1}$ M$_{\odot}$, $8\times10^{5}$ $h^{-1}$ M$_{\odot}$, and $4\times10^{5}$ $h^{-1}$ M$_{\odot}$, respectively.}
\tablenotetext{e}{Dark matter, gas, and stellar particle masses are $3\times10^{7}$ $h^{-1}$ M$_{\odot}$, $6\times10^{6}$ $h^{-1}$ M$_{\odot}$, and $3\times10^{6}$ $h^{-1}$ M$_{\odot}$, respectively.}
\end{deluxetable*} 

\subsection{Ionization and Thermal Structure}
\label{ionization and thermal}
Our basic SPH simulations, like most cosmological simulations to date, assume a UVB that is spatially uniform throughout the simulation volume and calculate the local ionization state of the gas assuming photoionization equilibrium with this background.
This approach misses the effects of self-shielding: where the optical depth to ionizing photons from the background is of order unity or more, the gas would in reality be exposed to an attenuated ionizing field.
This has consequences for both the ionization state and the thermal properties of the gas.
Indirectly, the gas dynamics is also affected by the modification of pressure forces.\\ \\
The omission of self-shielding implies that (in the absence of local sources) dense gas in the simulations sees an ionizing flux stronger than in reality.
As a result, the gas tends to be overionized. 
This is important for the \Lya~radiative transfer problem for two reasons.
First, the \Lya~emission mechanisms which seed the \Lya~photons depend not on the total gas density, but on the number densities of ions (\S \ref{emission processes}). 
Second, the transport of \Lya~photons depends on the \emph{neutral} hydrogen distribution, as only this ion provides scattering opacity (\S \ref{lya radiative transfer}).\\ \\
Self-shielding also has important effects on the thermal evolution of the gas.
As \cite{2001ApJ...562..605F} pointed out, neglecting self-shielding in a simulation with a prescribed uniform UVB introduces an artificially high rate of photoheating in dense regions and thus results in overestimated temperatures in these regions. 
Moreover, the presence of a penetrating ionizing background suppresses the cooling function \citep[e.g.,][]{1992MNRAS.256P..43E, 1996ApJS..105...19K, 1997ApJ...477....8W, 2005MNRAS.363....2K, 2009MNRAS.393...99W} and these effects could be amplified by the lack of a dynamical response of the gas to cooling, which would tend to make it denser and hence to cool even more rapidly. 
Although these effects are critical to accurately predicting the \Lya~cooling luminosity (\S \ref{lya emission}), previous \Lya~studies have either neglected them, made simplifying but not necessarily correct assumptions \citep[][]{2001ApJ...562..605F, 2010MNRAS.407..613G}, or have explored a range of prescriptions \citep[e.g.,][]{2005ApJ...622....7F, 2006ApJ...640..539Y}.\\ \\
We improve significantly over previous work by performing ionizing radiative transfer in post-processing to identify the self-shielded gas, rather than relying on simplified criteria (\S \ref{self shielding}). 
Since knowing the distribution of neutral hydrogen is a fundamental component of \Lya~radiative transfer, post-processing ionizing radiative transfer has previously been used by other groups for related problems \citep[e.g.,][]{2005ApJ...628...61C, 2009ApJ...696..853L, 2010ApJ...708.1048K, 2010ApJ...716..574Z}, but never before in focused studies of cooling emission. 
As we will show, the predicted \Lya~cooling luminosity is very sensitive to the state of the self-shielded gas. 
In order to obtain more robust predictions, we rerun our hydrodynamical simulations with the ionizing background turned off in regions above a certain density threshold (informed by our ionizing radiative transfer calculations) as an approximation to the consistent treatment of self-shielding.\\ \\ 
For the moment, we pause to discuss the star-forming gas, which also requires special treatment.

\begin{figure*}[ht] 
\begin{center} 
\includegraphics[width=1.0\textwidth]{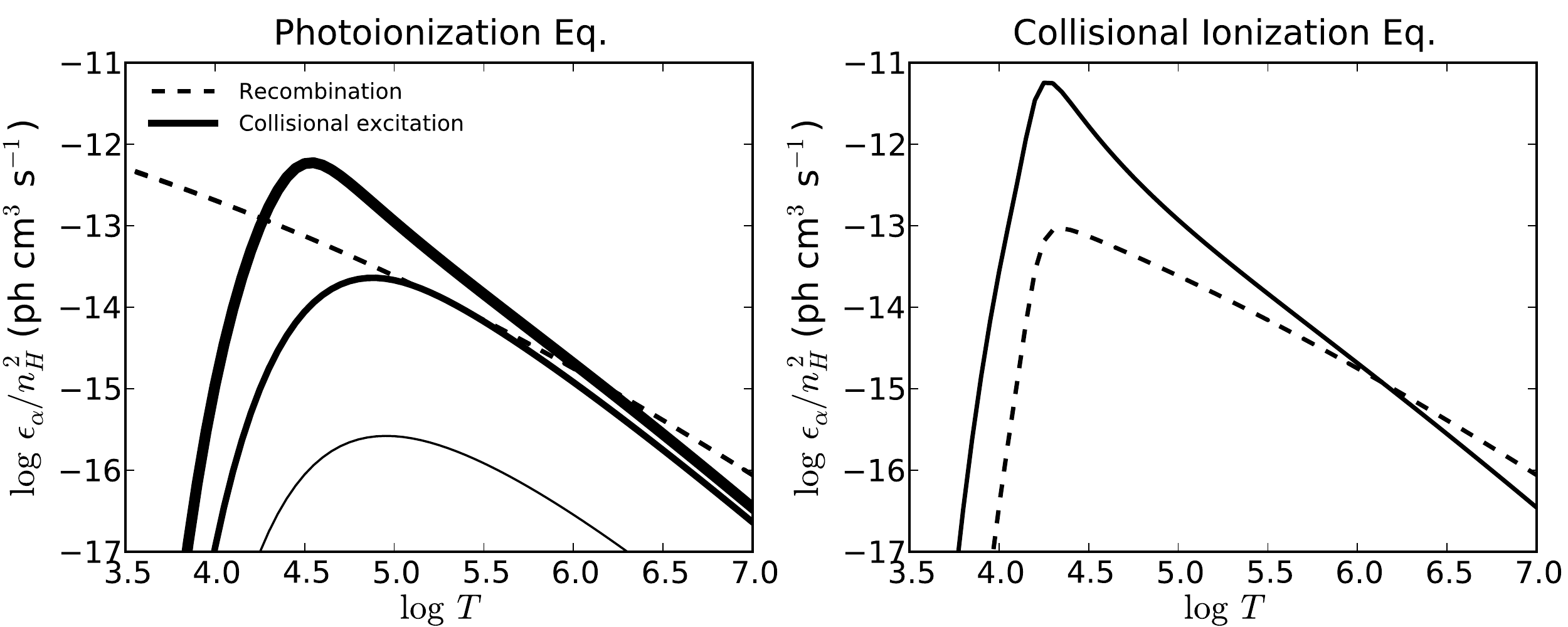} 
\end{center} 
\caption{\Lya~emissivity per $n_{\rm H}^{2}.$
\emph{Left:} Gas in ionization equilibrium with a hydrogen photoionization rate $\Gamma_{\rm HI}=10^{-12}$ s$^{-1}$. 
\emph{Right:} Gas in pure collisional ionization equilibrium.
The solid curves show the \Lya~emissivity from collisional excitation and the dashed curves show the emissivity from recombination.
For the photoionization equilibrium case, the collisional excitation curve depends on $n_{\rm H}$; we show $\log{(n_{\rm H}/{\rm cm}^{-3})}=-6,~-4,~{\rm and~-2}$ with curves of increasing thickness. 
The emissivity from collisional excitation scales with $n_{\rm HI} \propto \Gamma_{\rm HI}^{-1}$ in the photoionization equilibrium case. 
The unavoidable effects of collisional ionization are included in the photoionization equilibrium case. 
These curves assume that all the helium is in the form of HeIII.}
\label{emissivity} 
\end{figure*}

\subsection{The Multiphase ISM}
\label{multiphase ism}
The star-forming gas particles in the multiphase model carry effective ionic densities and temperatures that are mass-weighted averages of the hot and cold components \citep[][]{2003MNRAS.339..289S}.
While most of the mass in this model is in the $T=1,000$ K cold component,\footnote{The terminology with respect to temperature is somewhat discrepant in the contexts of galaxy formation and of the interstellar medium (ISM). While the $T=10^{4}$ K gas is termed cold in galaxy formation and throughout most of this paper, it is usually qualified as warm in the context of the ISM to distinguish it from the much cooler, star-forming molecular gas.} the hot component is in general much hotter ($T=10^{5}-10^{8}$ K). 
This results in high effective temperatures with simultaneously large neutral fractions, and therefore in high collisional excitation rates (\S \ref{collisional excitations}). 
As we will show, using the effective multiphase temperatures and ionic densities for the star-forming particles yields artificially high cooling luminosities, owing to the non-linearity of the emissivity function with respect to density and temperature. 
Moreover, in the multiphase model, supernovae are responsible for pressurizing the ISM and are therefore an additional source of energy.\\ \\
To obtain realistic results uncontaminated by feedback energy, it is necessary to exclude the star-forming particles from the cooling luminosity calculations. 
We explore two ways of doing this. 
First, we use the simulations with star formation, but ignore all the multiphase particles in the luminosity calculation. 
For this case, we assume that the star-forming particles are effectively optically thin for the purpose of transporting the \Lya~photons.
In reality, dust may destroy a portion of those photons, but we do not model this effect for this extreme prescription. 
This case approximates a multiphase medium in which cold neutral clumps embedded in a hot medium are either so compact that their covering factor is negligible, or in which they simply reflect the \Lya~photons and contribute only a small effective \Lya~optical depth \citep[e.g.,][]{1991ApJ...370L..85N, 2006MNRAS.367..979H}. 
A potential worry with this approach, however, is that it misses the cooling that occurs in particles with density above the star formation threshold. 
In our second approach, we make sure to capture all the cooling by using identical simulations but with the star formation model turned off, in which case we can simply sum over all the particles. 
The gas is allowed to become arbitrarily dense (as permitted by the resolution) and the absence of ionizing and mechnical feedback from embedded stars makes the density peaks very optically thick to \Lya~photons. 
By comparison with the case of optically thin star-forming gas, this prescription therefore allows us to also quantify the effects of the opacity provided by star-forming particles on the transport problem.\\ \\
We intentionally do not model the \Lya~photons produced by stars in this work in order to separate out the properties of pure cooling emission; we briefly discuss their importance in the discussion (\S \ref{discussion}) and in Appendix \ref{analytic considerations appendix}.

\section{Ly$\alpha$ EMISSION}
\label{lya emission}

\subsection{Emission Processes}
\label{emission processes}
\subsubsection{Recombination}
\label{recombinations}
Ionizing radiation (either from the cosmic background, local star formation, or an AGN) can photoionize gas that recombines and produces \Lya~photons.
Collisions in gas of sufficient density and temperature can also ionize hydrogen and be followed by the reemission of \Lya~photons via recombination.
We group these two processes in `recombination emission.'
Recombination emission produces \Lya~photons at a rate (in units of ph s$^{-1}$ cm$^{-3}$)
\begin{equation}
\label{recombinations eq}
\epsilon_{\alpha}^{\rm ph,rec} = f_{\alpha,{\rm rec}} \alpha_{\rm HI}^{\rm B}(T) n_{\rm HII} n_{e},
\end{equation}
where $\alpha_{\rm HI}^{\rm B}(T)\propto T^{-0.7}$ is the hydrogen case B recombination coefficient and $f_{\alpha,{\rm rec}}$ is the average number of \Lya~photons produced per case B recombination.
For gas at $T=10^{4}$ K that is optically thick to Lyman series transitions (so that higher-order Lyman series recombination photons can ultimately be degraded into a \Lya~photon), $f_{\alpha,{\rm rec}}=0.68$ \citep[][]{2006agna.book.....O}.
This fraction is only weakly dependent on temperature and so we assume this constant value throughout.

\begin{deluxetable*}{cll}[b]
\tablewidth{0pc}
\tablecaption{Prescriptions for Calculating the \Lya~Cooling Luminosity\label{prescriptions}}
\tabletypesize{\footnotesize}
\tablehead{\colhead{\#} & \colhead{Description} & \colhead{Notes}}
\startdata
1 & Standard hydro (uniform UVB and multiphase SF model), sum all particles & Overestimates \Lya~luminosity \\
   &                                                                                                       & ~~~due to UVB and multiphase model \\
2 & Standard hydro, with post-processing ionizing RT, no \Lya~from self-shielded gas & Satisfies energetic bound \\ 
3 & Standard hydro, with post-processing ionizing RT, CIE with $T_{\rm CIE}=10,000$ K & $\approx$ same \Lya~luminosity as 2\\
4 & Standard hydro, with post-processing ionizing RT, CIE with $T_{\rm CIE}=15,000$ K & $\approx10\times$ more \Lya~luminous than 3 \\
5 & Hydro with uniform UVB but no SF, sum all particles & Overestimates \Lya~luminosity \\
   &                                                                      & ~~~due to UVB only \\
\hline
6 & Self-shielding approx. for the UVB, with SF, sum all particles & Overestimates \Lya~luminosity \\
   &                                                                                                               & ~~~due to multiphase model only \\
7 & Self-shielding approx. for the UVB, no SF, sum all particles & Consistent \Lya~luminosity estimate \\
8 & Self-shielding approx. for the UVB, no SF, only sum $n_{\rm H}<0.13$ cm$^{-3}$ & Consistent \Lya~luminosity estimate \\
9 & Self-shielding approx. for the UVB, SF excluded in post-processing, sum all particles & Consistent \Lya~luminosity estimate
\enddata
\end{deluxetable*} 

\subsubsection{Collisional Excitation}
\label{collisional excitations}
Collisions can also excite the \Lya~line without ionizing hydrogen.
The \Lya~emissivity from this process is given by
\begin{equation}
\label{collisions eq}
\epsilon_{\alpha}^{\rm ph,coll} = C_{\rm Ly\alpha}(T)n_{\rm HI}n_{e},
\end{equation}
where $C_{\rm Ly\alpha}(T) \propto T^{-1/2} \exp{(-h\nu_{\alpha}/kT)}$ is the \Lya~collisional excitation coefficient in units of ph cm$^{3}$ s$^{-1}$.
Note that the \Lya~recombination and collisional excitation emissivities scale differently with temperature.
Moreover, whereas the recombination term is proportional to the HII number density, the collisional excitation term is proportional to the HI number density.
The relative importance of the two processes will therefore depend on the local temperature and ionization state of the gas.

\begin{figure*}[ht] 
\begin{center}
\mbox{
{\includegraphics[width=0.5\textwidth]{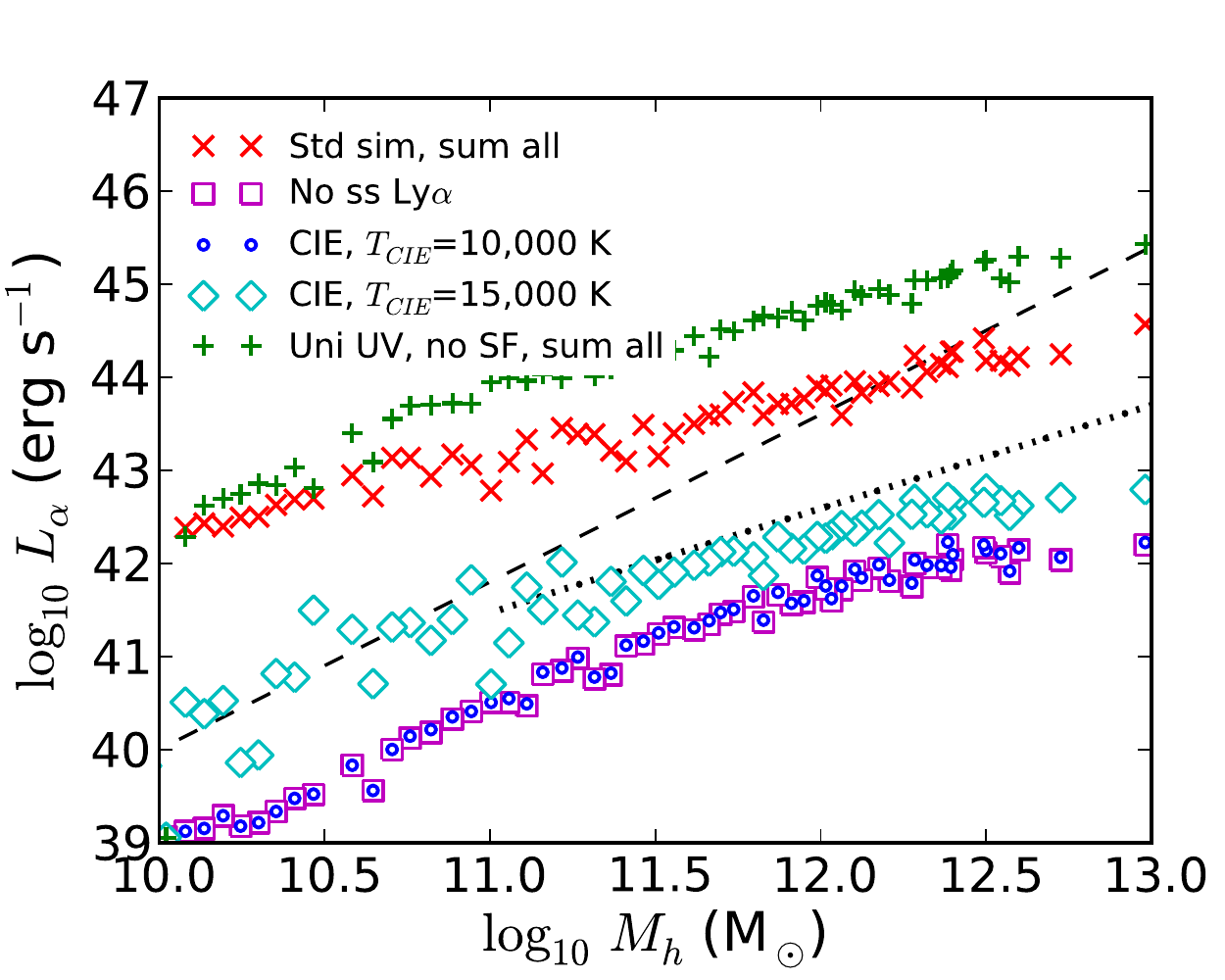}}
{\includegraphics[width=0.5\textwidth]{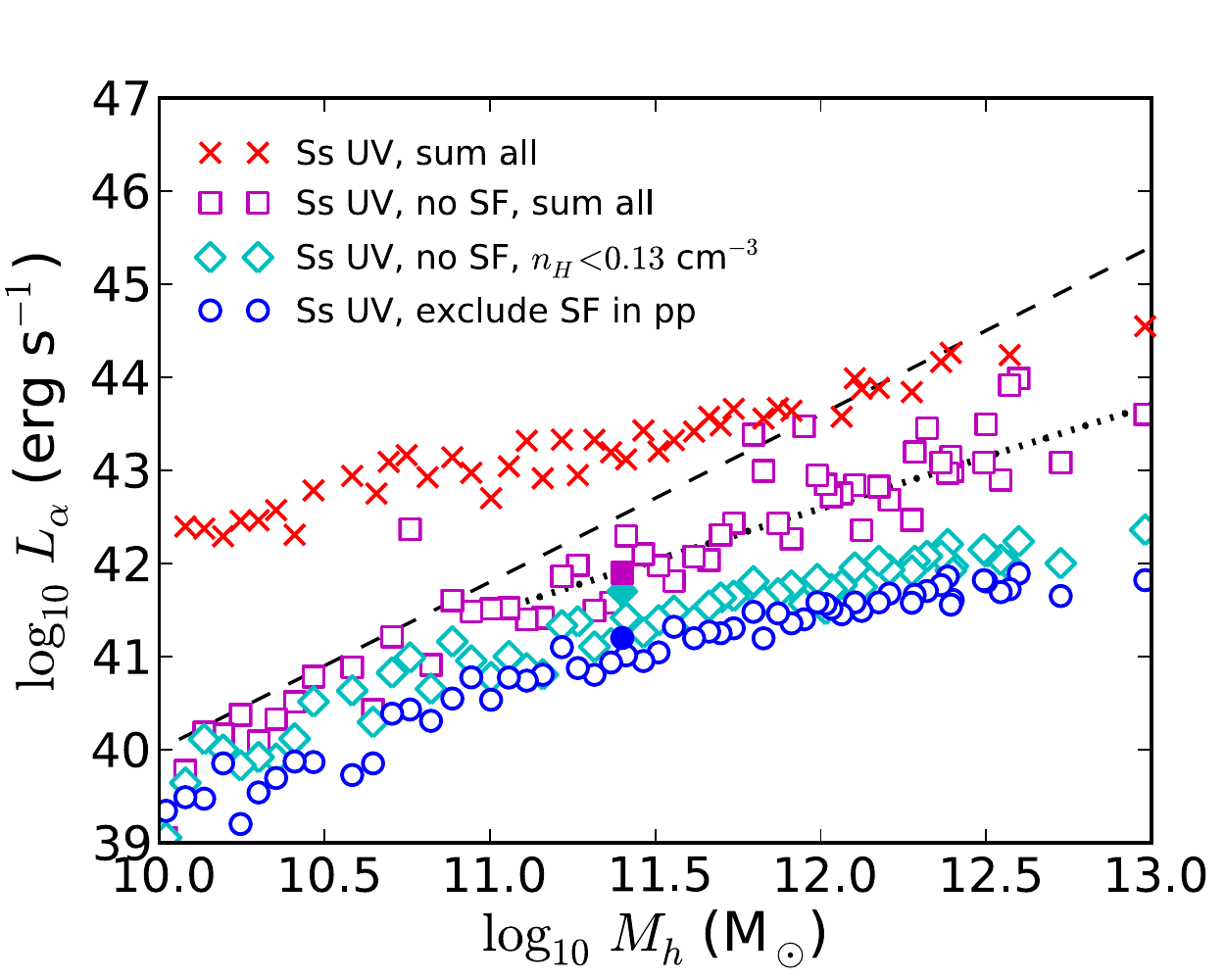}}
}\\
\end{center} 
\caption{\Lya~luminosity within the virial radius as a function of halo mass at $z=3$ calculated from our cosmological volume simulations, for different physical assumptions. 
Each point corresponds to a randomly selected halo. 
The different cases are defined in Table \ref{prescriptions}, where corresponding remarks are given.
\emph{Left:} Standard hydrodynamical simulation with uniform UVB and a multiphase star formation model, in some cases post-processed with the ionizing radiative transfer scheme to identify the self-shielded gas, corresponding to prescriptions 1, 2, 3, and 4. 
The green +s show the result of a simulation with a uniform UVB but no star formation, where the cooling luminosity is integrated over all the particles (prescription 5).
\emph{Right:} Simulations with the ionizing background turned off in regions where $n_{\rm H}>0.01$ cm$^{-3}$ to approximate self-shielding, corresponding to prescriptions 6, 7, 8, and 9. 
For the last three (most consistent) cases, the filled symbols show the results for the A1 halo in our zoom in simulation with 8$\times$ better mass resolution. 
Caution should be exercised when interpreting the quantitative details of the lowest-mass halos shown, as the hydrodynamics may not be fully converged (\S \ref{on the fly self shielding}).
The dashed lines show an analytic estimate for the maximum average cooling luminosity available from the release of gravitational potential energy (Appendix \ref{analytic considerations appendix}); the dotted lines show the more sophisticated analytic model of \cite{2009MNRAS.400.1109D} for their fiducial parameter $f_{\rm grav}=0.3$.}
\label{Lalpha vs M} 
\end{figure*}

\begin{figure*}[ht] 
\centering
\mbox{
{\includegraphics[width=0.50\textwidth]{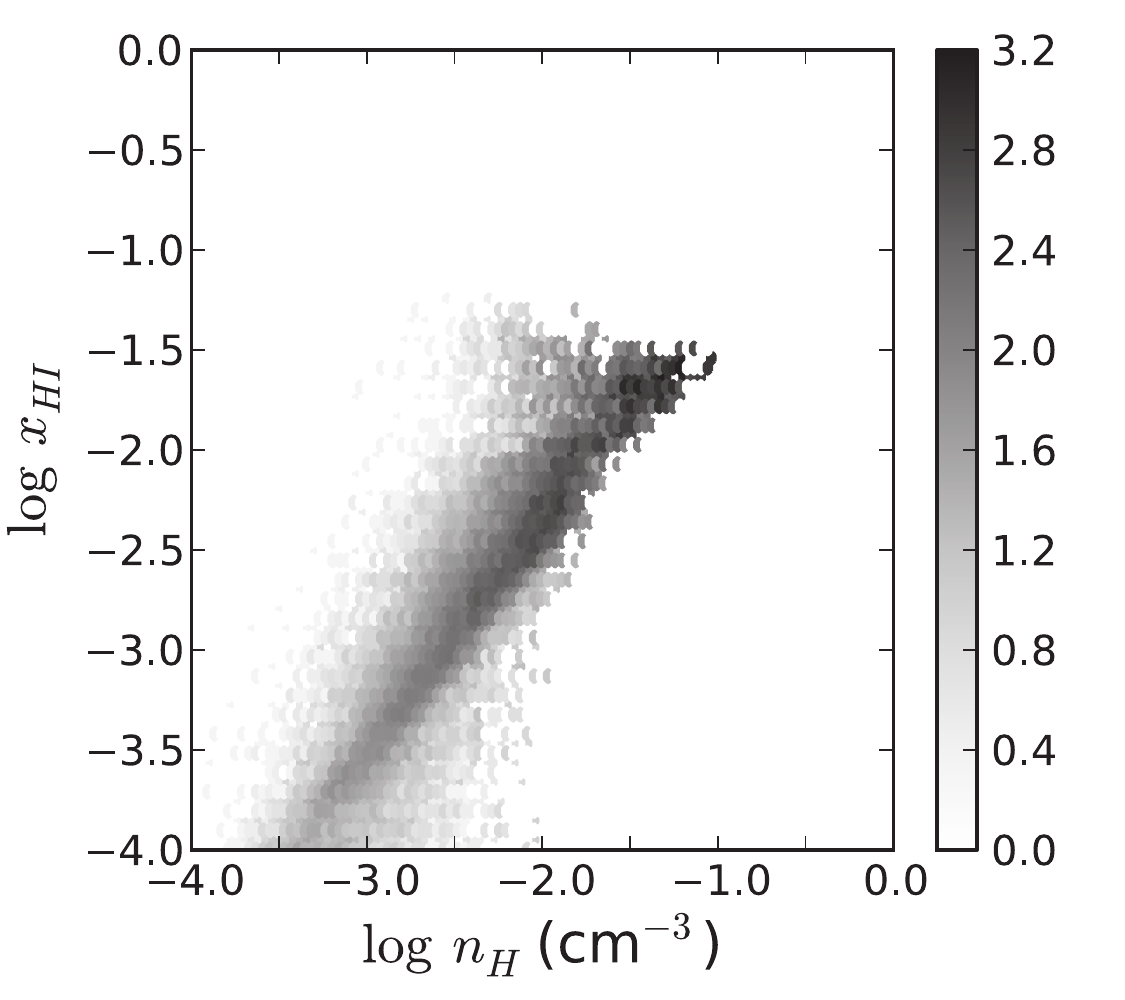}}
{\includegraphics[width=0.50\textwidth]{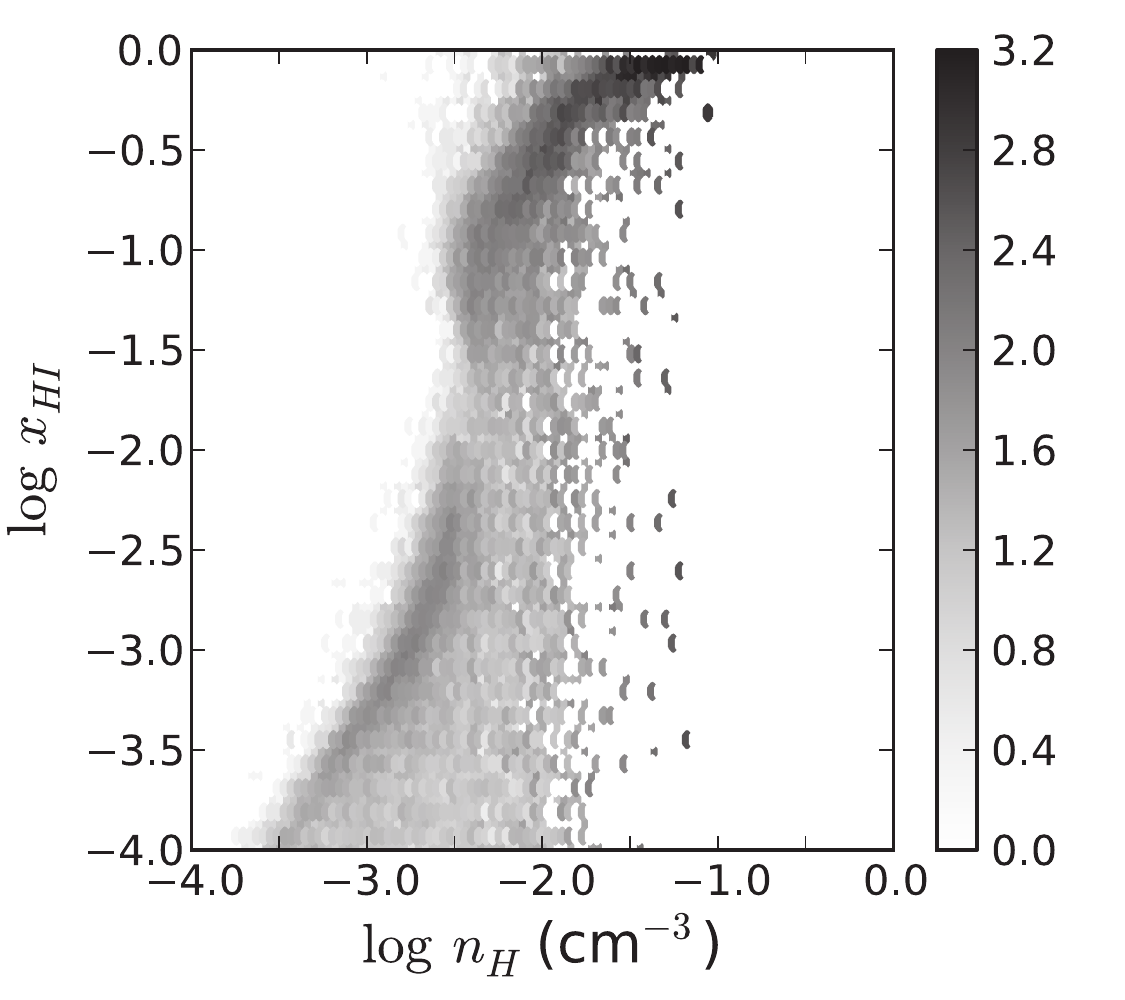}}
}
\caption{Hydrogen neutral fraction vs. total hydrogen number density in a cube of side length 1 comoving Mpc/h centered on the A1 system at $z=3$. \emph{Left:} Values for a standard simulation with a uniform UVB. \emph{Right:} Same quantity after post-processing with the ionizing radiative transfer method. 
The ionizing radiative transfer shows that the main effect of self-shielding is to create a vertical ``plume'' of neutral gas above a density $n_{\rm H}\sim0.01$ cm$^{-3}$. 
Figures \ref{A1 hydro vs physics} and \ref{T vs nH} illustrate the effects of self-shielding on the dynamics and thermal state of the gas, when it is approximated during the course of the hydrodynamical simulation. 
The 2D histograms are in arbitrary (but matching) logarithmic units and weighted by $n_{\rm H}^{2}$ to emphasize the regions where the two-body emission processes are most efficient.
Gas from multiphase, star-forming particles ($n_{\rm H}>0.13$ cm$^{-3}$) is excluded. 
}
\label{xHI vs nH} 
\end{figure*}

\subsubsection{Limiting Equilibrium Cases}
We assume ionization equilibrium, which is generally valid since the time scale $t_{\rm eq}\equiv[\Gamma_{\rm HI} + (\Gamma_{\rm HI,c}(T)+\alpha^{A}_{\rm HI}(T))n_{e}]^{-1}$ to reach equilibrium is small compared to the dynamical time scale in both optically thin and self-shielded gas.
The statistical equilibrium equation for hydrogen is
\begin{equation}
\label{ionization balance}
\Gamma_{\rm HI}n_{\rm HI} + \Gamma_{\rm HI,c}(T)n_{e}n_{\rm HI} = \alpha_{\rm HI}^{\rm A}(T) n_{e} n_{\rm HII},
\end{equation}
where $\Gamma_{\rm HI}$ is the photoionization rate, $\Gamma_{\rm HI,c}(T)$ is the collisional ionization coefficient, and $\alpha_{\rm HI}^{\rm A}(T)$ is the case A recombination coefficient.\\ \\
Physical intuition can be gained by considering the two limiting cases of photoionization equilibrium and of pure collisional ionization equilibrium (CIE; $\Gamma_{\rm HI} \ll \Gamma_{\rm HI,c}(T)n_{e}$). 
As we will show, the two cases are directly relevant to our problem: 
While most of the cosmic volume is well approximated by the photoionization equilibrium regime, the dense cold gas (including the cold streams of interest) can self-shield from the external ionizing radiation and is then more accurately described by the the pure CIE case. 
Figure \ref{emissivity} shows $\epsilon_{\alpha}^{\rm ph}/n_{\rm H}^{2}$ from both recombination and collisional excitation.
The left panel shows the case of gas in ionization equilibrium with a photoionization rate $\Gamma_{\rm HI}=10^{-12}$ s$^{-1}$ \citep[approximately the magnitude of the cosmic ionizing background at $z\approx2-4$, e.g.][]{2008ApJ...682L...9F, 2008ApJ...688...85F}, and the right panel shows the case of gas in collisional ionization equilibrium.
The most important point to note is that the collisional excitation contribution is exponentially sensitive to temperature, at the temperatures $T\sim 10^{4}$ K characteristic of cold accretion streams, in both cases. 
As collisional excitation is the dominant \Lya~cooling mechanism, it is critical to accurately capture the thermal state of the emitting gas. 
The curves shown in Figure \ref{emissivity} assume that all the helium is in the form of HeIII for simplicity. 
The emissivity from both processes is only weakly sensitive to the helium ionized fractions, except for temperatures $T\lesssim10^{4}$ K for the CIE case and at very high densities in the photoionization case, when most of the hydrogen is neutral and helium dominates the free electrons. Since little \Lya~emission originates from these regimes, our results are broadly insensitive to the helium ionization state, although we do solve for the correct helium ionization fraction in self-shielded regions in our simulations with the on-the-fly self-shielding approximation (\S \ref{on the fly self shielding}).\\ \\
In our simulations, the \Lya~luminosity is evaluated directly from the SPH particles. 
Specifically, each particle is assigned a \Lya~luminosity $L_{\alpha,{\rm p}}^{\rm ph} \equiv V_{\rm p} (\epsilon_{\alpha}^{\rm ph,rec} + \epsilon_{\alpha}^{\rm ph,col})$, where the emissivity terms are evaluated using its density, ionization state, and temperature, and $V_{\rm p} \equiv M_{\rm p}/\rho_{\rm p}$ is its volume, defined as the ratio of its mass to its density. 
This approach is desirable as it accurately takes into account the clumping of the gas on small scales, relevant to calculate the emission from the density-squared processes, and because it avoids artificial mixing that could occur if hot and cold phases were averaged in a gridding procedure. 
Such artificial mixing could boost the predicted \Lya~luminosity by a large factor owing to the non-linearity of the emissivity function. 

\begin{figure*}[ht] 
\begin{center} 
\mbox{
{\includegraphics[width=1.0\textwidth]{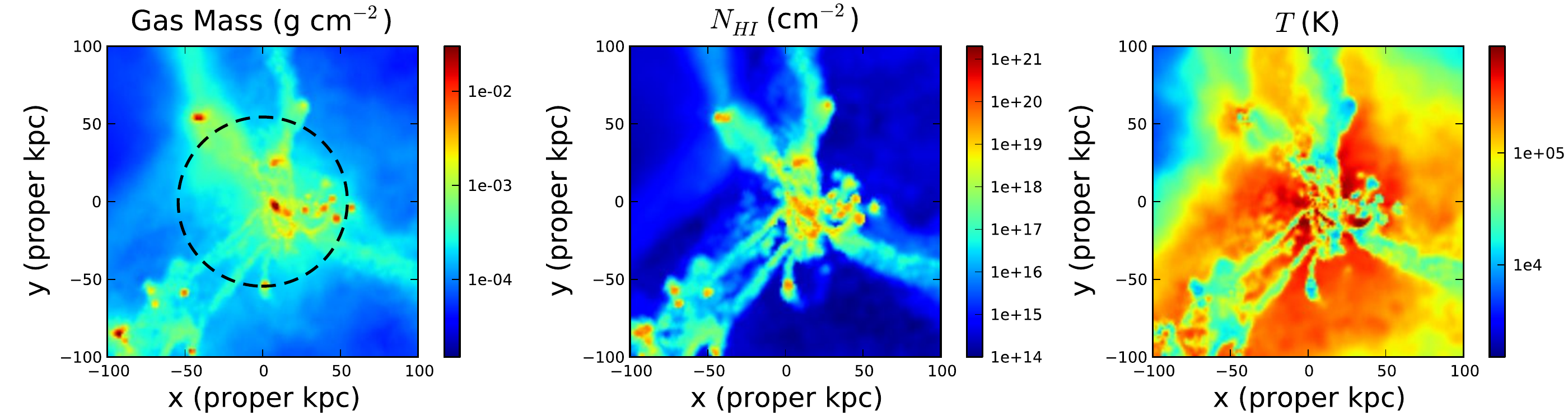}}
}\\
\mbox{
{\includegraphics[width=1.0\textwidth]{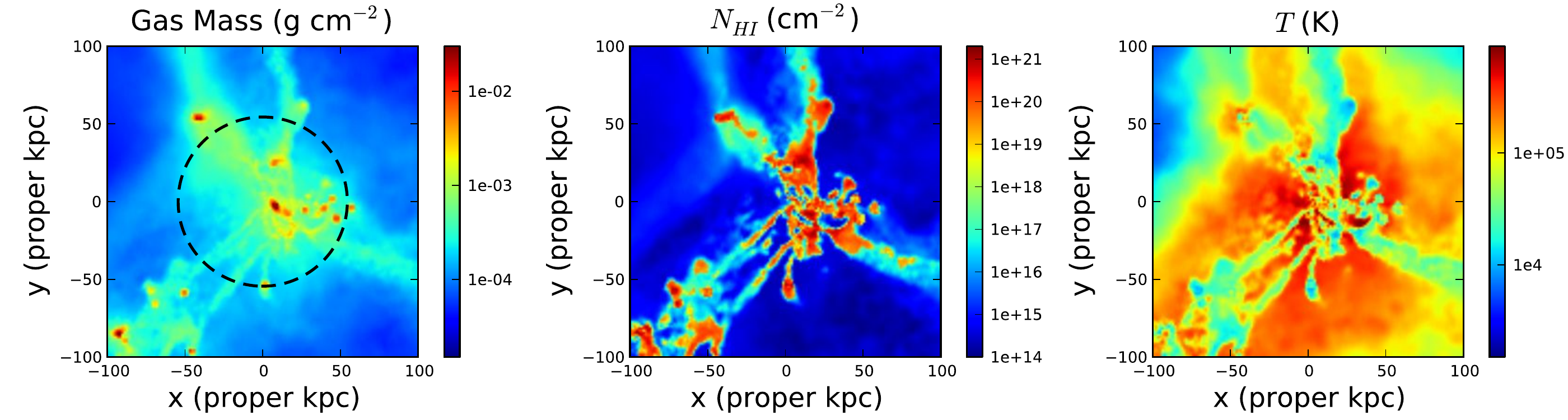}}
}\\
\mbox{
{\includegraphics[width=1.0\textwidth]{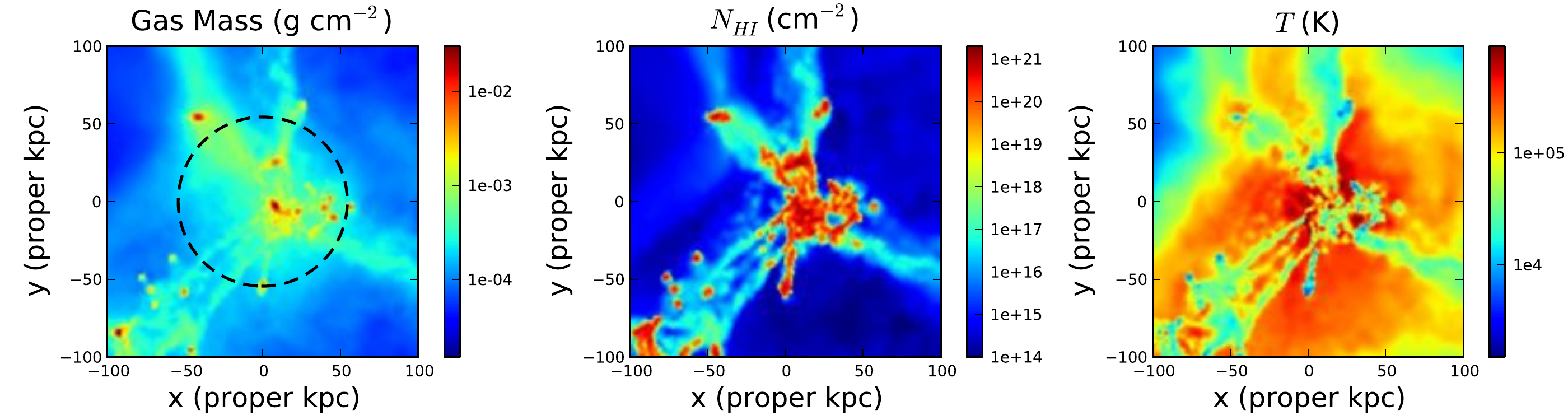}}
}
\end{center} 
\caption{Hydrodynamical properties of the A1 system at $z=3$ as a function of the self-shielding treatment. 
The left column shows the total projected gas mass, the central column shows the neutral hydrogen column density, and the right column shows the projected gas temperature. 
In all cases, the projected depth is 1 comoving Mpc/h. 
The projected temperature is weighted by density squared to emphasize the dense filaments and the virial radius of the halo is indicated by the dashed circles.
\emph{Top:} Standard simulation output, with a uniform ionizing background. 
\emph{Middle:} Same, but post-processed with the ionizing radiative transfer scheme to identify the self-shielded regions (\S \ref{post processing self shielding}). 
\emph{Bottom:} Same initial conditions but with the on-the-fly self-shielding approximation, in which the ionizing background is turned off in regions with $n_{\rm H}>0.01$ cm$^{-3}$ as the simulation proceeds to capture the effects on the thermal and dynamical evolution of the gas. 
Figure \ref{T vs nH} illustrates the effects of self-shielding on the temperature structure of the gas more explicitly. 
The simulations shown include star formation.}
\label{A1 hydro vs physics} 
\end{figure*}

\begin{figure*}[ht] 
\centering
\mbox{
{\includegraphics[width=0.50\textwidth]{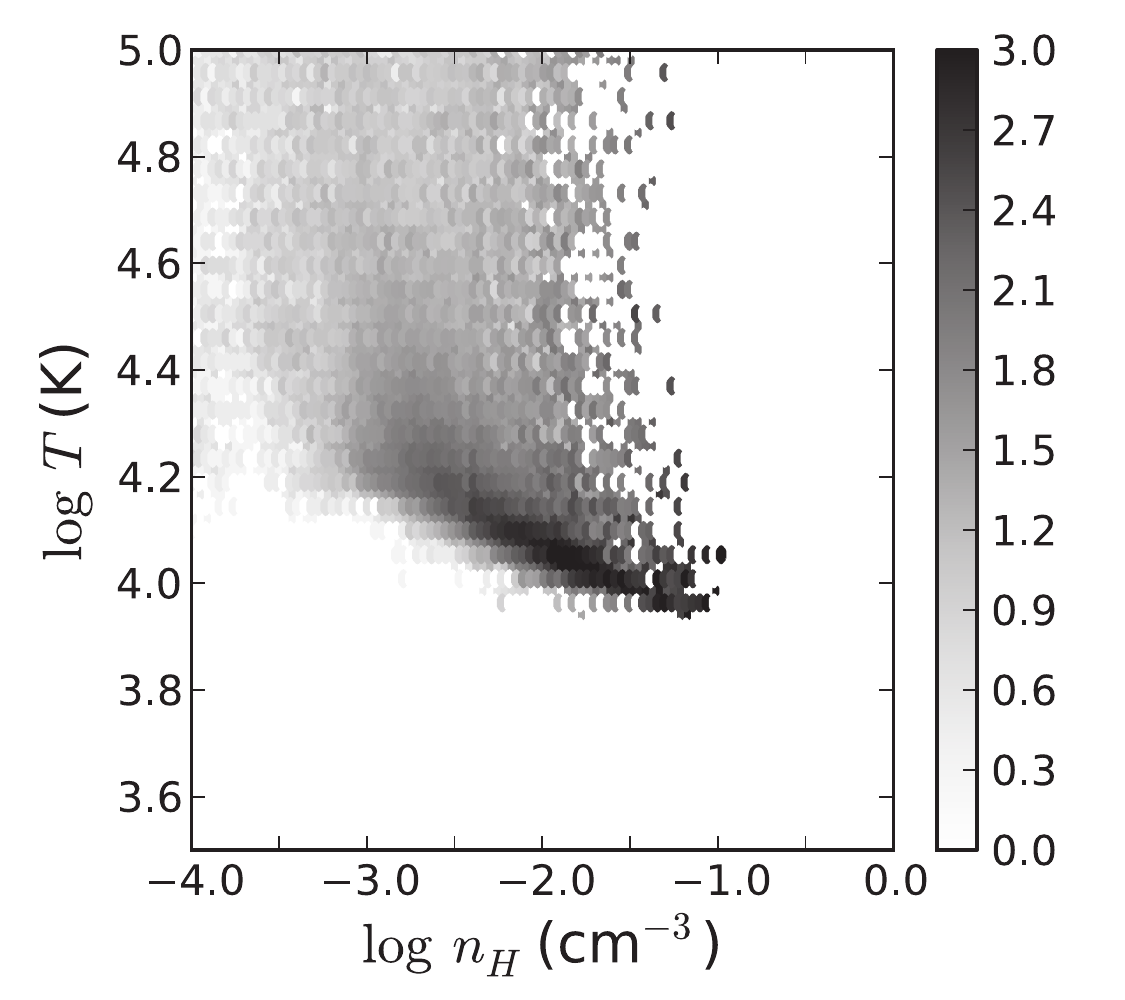}}
{\includegraphics[width=0.50\textwidth]{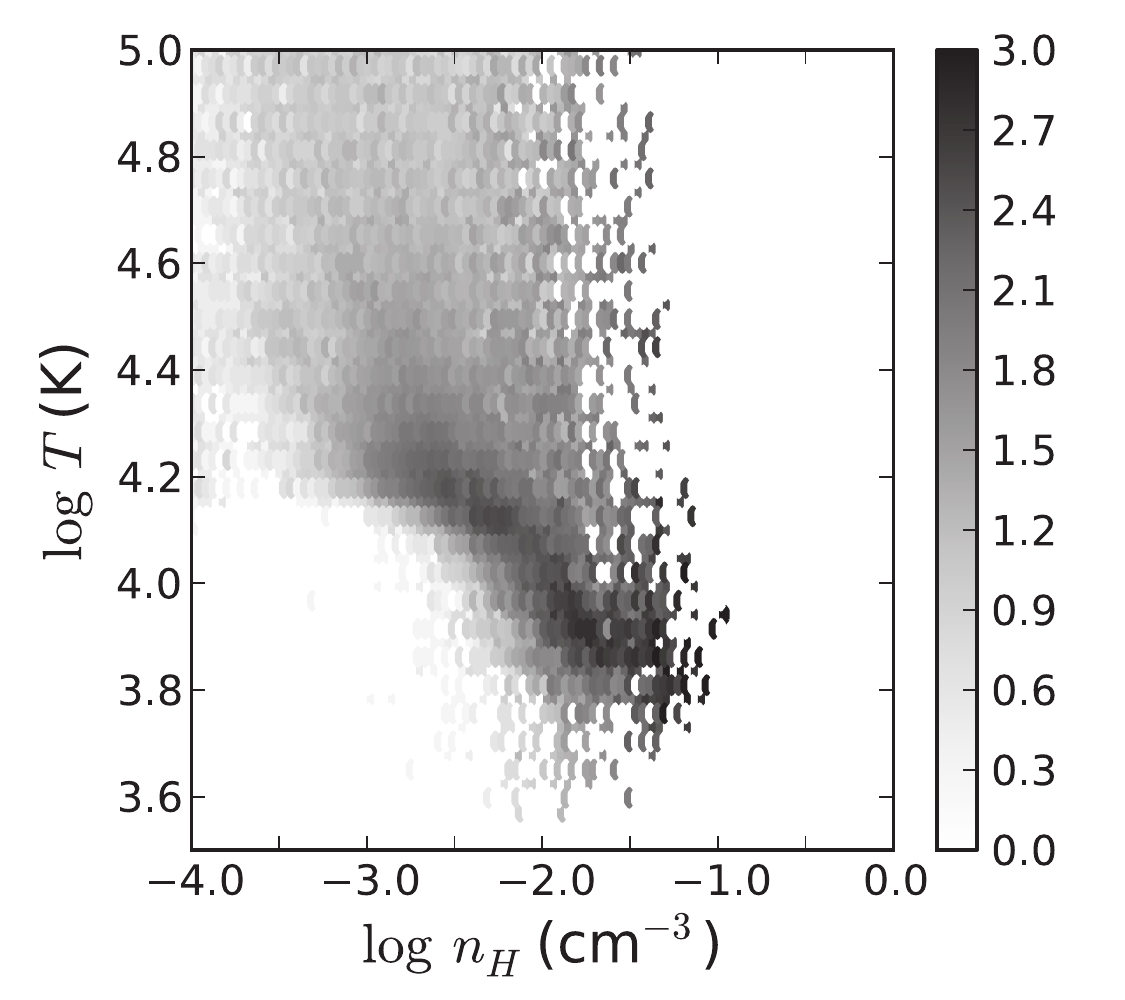}}
}
\caption{Gas temperature vs. total hydrogen number density in a cube of side length 1 comoving Mpc/h centered on the A1 system at $z=3$. \emph{Left:} Values for a standard simulation with a uniform UVB (top panel of Fig. \ref{A1 hydro vs physics}). These also apply for the simulation post-processed with the ionizing radiative transfer scheme, since it does not update the temperatures (middle panel of Fig. \ref{A1 hydro vs physics}). \emph{Right:} Same quantity, but for the simulation with the ionizing background turned off in regions with $n_{\rm H}>0.01$ cm$^{-2}$ during the course of the hydrodynamical calculation as an approximation to the effects of self-shielding (bottom panel of Fig. \ref{A1 hydro vs physics}). 
The self-shielded gas is generally cooler (with $T\lesssim10^{4}$ K) when its evolution is consistently modeled as a result of the suppression of artificial photoheating and the enhancement of its cooling function. 
The 2D histograms are in arbitrary (but matching) logarithmic units and weighted by $n_{\rm H}^{2}$ to emphasize the regions where the two-body emission processes are most efficient.
Gas from multiphase, star-forming particles ($n_{\rm H}>0.13$ cm$^{-3}$) is excluded.}
\label{T vs nH} 
\end{figure*}

\subsection{Self-Shielding}
\label{self shielding}
Since the emission processes scale with density squared (eqs \ref{recombinations eq}-\ref{collisions eq}), the emissivity peaks in the densest regions, which are the most likely to self-shield. 
Because our hydrodynamical simulations lack proper ionizing radiative transfer, they do not correctly capture self-shielding (\S \ref{ionization and thermal}). 
As explained in \S \ref{multiphase ism}, naively integrating over multiphase SPH particles could also induce large errors in the predicted cooling luminosity. 
As we will show, \emph{it is necessary to both exclude multiphase particles from the calculation and to model self-shielding} to accurately predict the cooling luminosity. 
The rest of this section is dedicated to demonstrating the importance of each potential source of error and to developing a consistent approximation to the \Lya~cooling luminosity.\\ \\
We follow the following steps:
\begin{enumerate}
\item Naively calculate the \Lya~luminosity from a simulation with standard UV background and star formation treatments.
\item Using post-processing ionizing radiative transfer, identify the self-shielded gas in step 1.
\item Using the post-processed output, illustrate how the predicted \Lya~luminosity depends on the assumed thermal state of the self-shielded gas. 
\item Rerun a hydrodynamical simulation with the same initial conditions, but with the ionizing background turned off in self-shielded regions on the fly as an approximation to the self-consistent effects of self-shielding.
\item By rerunning identical simulations with star formation turned off, separate the effects of incorrectly including multiphase particles from those of ignoring self-shielding.
\end{enumerate}

\subsubsection{Post-Processing Self-Shielding}
\label{post processing self shielding}
A technical description of our ionizing radiative transfer code is provided in Appendix \ref{ionizing radiative transfer}.
Briefly, the hydrogen photoionization rate of the cosmic background, $\Gamma_{\rm HI}^{\rm bkg}$, is specified and taken as the boundary condition at the faces of the cubical radiative transfer volume. 
For the radiative transfer calculations, the simulation outputs are interpolated onto a Cartesian grid taking into account the smoothing kernels, with $N_{\rm p}$ grid points along each dimension. 
We employ the fiducial choice $N_{\rm p}=256$ and a radiative transfer volume of (1 comoving Mpc/h)$^{3}$, centered around each halo considered, which convergence tests suggest is sufficient (\S \ref{convergence section}). 
Rays normal to each of the six faces are then sent inward and the optical depth to ionizing photons is calculated along each ray.
Given the attenuated photoionization rate at each point, the ionization equilibrium is updated taking into account photoionization, collisional ionization, and recombination.
The procedure is iterated until the ionized fraction has converged in all the cells. 
In solving for the equilibrium ionization balance, the gas temperatures used are those provided by the hydrodynamical simulation.
These should be accurate in the optically thin regions and therefore our scheme should accurately capture the onset of self-shielding.
In this post-processing treatment, the temperature structure will however be inaccurate in the self-shielded regions, since the modifications of the heating and cooling functions are not properly modeled in the hydrodynamical calculation. 
As outlined above, we address this in two ways: first, we explore a range of prescriptions for the self-shielded gas, illustrating the sensitivity of the predictions to these prescriptions; we then subsequently improve the accuracy of our calculations by approximating the self-consistent thermal evolution of the gas with simulations in which the ionizing background is switched off in dense regions.  
The self-shielded cells are defined as those that see an attenuated ionizing background, $\langle e^{-\tau_{i}}\rangle<\langle e^{-\tau_{i}}\rangle _{\rm crit}$ with $\langle e^{-\tau_{i}}\rangle _{\rm crit}\equiv 0.1$, where $\langle e^{-\tau_{i}}\rangle$ is the angle-averaged attenuation factor in the cell after post-processing. Since the optical depth rapidly increases within a self-shielded region, the results are weakly sensitive to the choice of the self-shielding threshold.\\ \\
To investigate how the \Lya~luminosity depends on assumptions regarding the self-shielded gas (steps 1, 2, and 3 above), we start with hydrodynamical simulations with standard treatments of the UV background and of star formation. 
Halos covering a broad range of masses are selected from our cosmological volume simulation gdm40n512 at $z=3$ and the luminosity of each is defined as the sum of the luminosities of the gas particles contained within its virial radius. 
The same prescriptions are also applied to the A1 halo at $z=3$, which will be used for the \Lya~radiative transfer calculations (\S \ref{lya radiative transfer}). 
We explore three cases that are illustrative of the range of possibilities \citep[for similar prescriptions, see][]{2005ApJ...622....7F}:\\ \\
{\bf Naively sum all the particles.} In this simplistic prescription, we simply sum all the SPH particles within the virial radius. 
This includes dense particles that would in reality self-shield but that are artificially illuminated by a uniform ionizing background, and star-forming particles that carry effective multiphase values for their temperature and ionization state, which will introduce luminosity errors as described in \S \ref{multiphase ism}.\\ \\
{\bf No \Lya~emission from self-shielded gas.} Self-shielded gas may become neutral and cool substantially.
If the gas is not photoionized and cools below $T\approx10^{4}$ K, its \Lya~emissivity is severely suppressed (Fig. \ref{emissivity}).
To illustrate how this case might differ from the naive calculation above, we model it by the extreme assumption that self-shielded gas does not produce \Lya~photons at all. 
To transport the \Lya~photons, we assume that this gas has a temperature $T=10,000$ K.\\ \\
{\bf Collisional ionization equilibrium.} Somewhat intermediate between the two above cases, self-shielded gas could settle to collisional ionization equilibrium (CIE) with a temperature $T_{\rm CIE}\gtrsim10,000$ K if gravitational heating is sufficiently efficient.
In this case, the neutral hydrogen fraction is given by
\begin{equation}
x_{\rm HI} = 
\frac{1}{1 + \Gamma_{\rm HI,c}(T_{\rm CIE})/\alpha_{\rm HI}^{\rm A}(T_{\rm CIE})}
\end{equation}
and the \Lya~emissivity $\epsilon_{\alpha}/n_{\rm H}^{2}$ is a well defined function of temperature, as shown in the right panel of Figure \ref{emissivity}. 
We explore how the luminosity depends on the prescribed CIE temperature, for $T_{\rm CIE}=10,000$ K and 15,000 K. 
As we will see in our simulations that approximate the effects of self-shielding, the self-shielded gas cools very effectively, so that higher temperatures are not expected. 
Furthermore, the proximity of 15,000 K to the peak of the cooling curve implies that this case is already quite optimistic.\\ \\
For the latter two cases, in which we assume either no emission from self-shielded gas or CIE, we exclude emission from the star-forming particles that might fall outside of the self-shielded regions, in order to avoid potential confusion with artificially high luminosities from multiphase particles. 
Table \ref{prescriptions} summarizes the different prescriptions explored for calculating the \Lya~cooling luminosity; the above cases are labeled 1$-$4.\\ \\
The left panel of Figure \ref{Lalpha vs M} shows how the \Lya~cooling luminosity varies with halo mass at the fiducial redshift $z=3$ for these different prescriptions. 
The dashed curve shows an analytic estimate of the average maximum cooling luminosity achievable from the release of gravitational potential energy as a function of halo mass (see Appendix \ref{analytic considerations appendix}; also Goerdt et al. 2010\nocite{2010MNRAS.407..613G} for similar ideas). 
Briefly, the $\Lambda$CDM cosmology predicts the average mass accretion rate onto dark matter halos as a function of mass and redshift \citep[e.g.,][]{2008MNRAS.383..615N, 2009MNRAS.398.1858M, 2010arXiv1001.2304F}, as well as the shape of the dark matter halo potential wells \citep[e.g.,][]{1990ApJ...356..359H,1997ApJ...490..493N}. 
The product of the halo potential well depth with the gas mass accretion rate provides an estimate of the rate at which gravitational potential energy that can be radiated is ``injected'' into the halo. 
Figure \ref{Lalpha vs M} assumes an optimistically high efficiency factor $f_{\rm eff}=1$ (eq. \ref{Egrav max}). 
As we discuss in Appendix \ref{Lalpha vs M}, this gravitational power estimate is not strictly an upper bound for the total amount of cooling achievable even in the case of pure accretion, without feedback from stars and AGN, since the accretion streams are embedded in the cosmic ionizing background, which can transfer further energy to them. 
This contribution is however included in our simulations and circumstantial evidence suggests that it does not dominate. In fact, in our most realistic simulations with a consistent self-shielding approximation -- shown in the right panel of Figure \ref{Lalpha vs M} and to be discussed below -- the gravitational power upper bound is never systematically violated, and a simulation of our A1 halo with the ionizing background completely turned off yields a nearly equal cooling luminosity (within 30\%) at $z=3$  as our prescription 9 when excluding the multiphase star forming regions in the same manner. \\ \\
The more sophisticated analytic model of \cite{2009MNRAS.400.1109D}, shown by the dashed lines for their fiducial efficiency parameter $f_{\rm grav}=0.3$, includes a factor $f_{\rm cold}(M)$ accounting for the decreasing fraction of cold gas in massive halos \citep[e.g.,][]{2005MNRAS.363....2K, 2009MNRAS.395..160K}. 
This model therefore predicts lower cooling luminosities than the above upper bound, with a shallower mass dependence at large masses that is in better agreement with the simulation data points. 
It is important to note here that the cooling luminosities shown in Figure \ref{Lalpha vs M} are ``theoretical'' or ``intrinsic'', meaning that they include all the photons emitted within the virial radii of the halos. 
These luminosities will in general be higher than the observationally inferred luminosities, which only include the emission above a certain surface brightness threshold determined by the observation. 
Furthermore, a certain fraction of the emitted photons are in practice absorbed by the intervening intergalactic medium (IGM). 
The \cite{2009MNRAS.400.1109D} data points plotted here (which were computed for a 50\% IGM transmission factor by these authors) have been multiplied by a factor of 2 for a fair comparison with our simulation data points (which assume 100\% transmission). 
In future work, we will quantify how the predicted theoretical luminosities translate into observational ones.\\ \\
There are two main points to take away from the left panel of Figure \ref{Lalpha vs M}. 
First, the predicted \Lya~cooling luminosity is extremely sensitive to the treatment of the dense gas, with the range exceeding three orders of magnitude at $M_{h}\sim10^{10}$ M$_{\odot}$. 
Second, some prescriptions actually lead to unphysical results, as comparison with the analytic upper bound indicates that they emit more \Lya~power than is available from the release of gravitational energy by orders of magnitude (in Appendix \ref{analytic considerations appendix}, we show that photoionization from the cosmic background cannot physically account for such large luminosities either). 
This is the case, in particular, for a standard simulation with optically thin ionizing balance and a multiphase star formation model in which all the gas particles are naively summed over (red x's).
Unsurprisingly, the cases of no \Lya~emission from self-shielded gas (magenta squares) and of CIE with $T_{\rm CIE}=10,000$ K (blue circles) yield nearly identical luminosities, since the cooling curve is already strongly suppressed at this temperature. 
The case of CIE with $T_{\rm CIE}=15,000$ (cyan diamonds) yields more optimistic \Lya~luminosities, although these push the limit of the power that can be provided by the release of gravitational potential energy alone (the dashed line in Fig. \ref{Lalpha vs M}) at low masses. 
Prescription number 5 (green +'s), which uses a simulation in which the multiphase star formation model was turned off, but with a uniform ionizing background penetrating deep into the dense gas, illustrates that artificial photoheating alone can boost the cooling luminosities by orders of magnitude.\\ \\
The critical question is therefore: What is the correct \Lya~cooling luminosity of the cold streams? 
To address this, we develop a simple approximation to the self-consistent evolution of the gas properties with self-shielding. 
When self-shielding is properly modeled, it will also be possible to show that naively integrating over multiphase particles alone can also artificially boost the cooling luminosity by a large factor.

\subsubsection{On-the-Fly Self-Shielding}
\label{on the fly self shielding}
In Figure \ref{xHI vs nH} we plot the hydrogen neutral fraction $x_{\rm HI} \equiv n_{\rm HI} / n_{\rm H}$ as a function of total proper hydrogen number density $n_{\rm H}$ for the gas around the A1 halo at $z=3$.
The panel on the left shows this distribution for the standard simulation with a uniform ionizing background. 
The panel on the right shows exactly the same quantity after the post-processing ionizing radiative transfer. 
The effect of self-shielding is clear.
Roughly, it generates a vertical ``plume'' above $n_{\rm H}\sim0.01$ cm$^{-3}$, indicating the fact that the gas becomes mostly neutral above this density. 
This motivates our approximation to the self-consistent evolution of self-shielded gas in the hydrodynamical simulations.
Namely, we rerun simulations with exactly the same initial conditions and other physical parameters, but set the ionizing background to zero in regions where the density exceeds the fiducial threshold $n_{\rm H} = 0.01$ cm$^{-3}$ \citep[for an alternative scheme in which the UVB is turned off where the gas is optically thick to ionizing photons on a scale $\sim0.1-1$ kpc, see][]{2006ApJ...644L...1S, 2007ApJ...657L..69L, 2009ApJ...704.1640L}.
By turning the ionizing background off in dense regions on the fly, their thermal and dynamical properties are consistently evolved with the modified cooling and heating functions, and the corresponding dynamical response. 
The ``CDB'' simulation analyzed by \cite{2010MNRAS.407..613G} employed an analogous scheme, but with a density threshold 10$\times$ higher, $n_{\rm H} = 0.1$ cm$^{-3}$; in \S \ref{discussion}, we argue that this difference likely explains much of the discrepancy with our results. 
Some uncertainty is introduced by our choice of a fixed density threshold for self-shielding, and in the future it would be useful to improve the methodology by performing proper ionizing radiative transfer on the fly, which our codes do not allow us to do at present. 
There are however reasons to believe that this choice is a good one, which we outline next.\\ \\
Figure \ref{A1 hydro vs physics} summarizes the hydrodynamical properties (total gas distribution, neutral gas distribution, and temperature structure) for the A1 system at $z=3$ for the different treatments of self-shielding: standard uniform ionizing background, post-processing ionizing radiative transfer, and the on-the-fly self-shielding approximation. When the ionizing radiative transfer is taken into account, the neutral hydrogen column density of the cold streams can be greatly enhanced, especially in the higher density regions close to the central and satellite galaxies, indicating the fact that they self-shield (some of the cold gas at larger radii however remains optically thin). 
The ionization structure obtained with the on-the-fly self-shielding approximation is furthermore remarkably similar to the one obtained with the post-processing ray tracing scheme, supporting the validity of using our simple density criterion during the course of the hydrodynamical simulation. 
The simulation with on-the-fly self-shielding is the most accurate as it consistently captures the thermal evolution and dynamical response of the self-shielded gas. 
Figure \ref{T vs nH} illustrates the effects of self-shielding on the temperature structure of the gas more explicitly: the self-shielded gas with $n_{\rm}>0.01$ cm$^{-3}$ is generally cooler (with $T\lesssim10^{4}$ K) when its evolution is consistently modeled. 
This simply results from the suppression of artificial photoheating and the enhancement of the cooling function in CIE. 
At these temperatures, the gas radiates very inefficiently in \Lya, which provides further evidence that the simple self-shielding density threshold is not introducing large errors: the \Lya~cooling luminosity versus halo mass predicted using prescription 9 (discussed below) is quite close to what is obtained by effectively suppressing the \Lya~emission from all the self-shielded gas, as in prescriptions 2 and 3 in which the self-shielded gas is identified using a ray tracing method and does not assume a particular density threshold. We have also run a simulation of the A1 halo with the ionizing background completely turned off, so that all the cooling in this case originates from gravitational energy and requires no self-shielding correction. The cooling luminosity for this simulation equals the one obtained with the simulation with on-the-fly self-shielding within $\sim$30\%, when the star-forming regions are identically excised. 
We are therefore confident that our simple density threshold for self-shielding yields relatively accurate results. 
\\ \\
The prescriptions for calculating the \Lya~cooling luminosity from the simulations with on-the-fly self-shielding approximation are labeled 6$-$9 in Table \ref{prescriptions} and the corresponding results are shown in the right panel of Figure \ref{Lalpha vs M}. 
For the last three (most consistent) cases, the filled symbols show the results for the A1 halo in our zoom in simulation with 8$\times$ better mass resolution. 
Prescription 6 (red x's), in which self-shielding is modeled but in which we naively sum over multiphase particles, demonstrates how the multiphase particles alone can produce artificially high cooling luminosities; these should therefore always be excluded, or treated separately. 
As an aside, comparison of prescriptions 1, 5, and 6 indicates that having the high density regions turn into multiphase particles limits the amount of artificial photoheating by effectively shielding the very dense gas. 
Three prescriptions (7, 8, and 9) correspond to physically plausible cases: one with the on-the-fly self-shielding approximation and star formation, but with multiphase star-forming particles excluded from the \Lya~luminosity sum (9; blue circles); one also with the on-the-fly self-shielding approximation, but with the star formation model turned off, summed over all the particles (7; magenta squares); and the intermediate case of summing only the particles with $n_{\rm H}<0.13$ cm$^{-3}$ at which the gas would have become multiphase if the star formation model had been on (cyan diamonds; 8). 
All three are physically realistic in the sense that they are uncontaminated by either artificial photoheating or by the sub-resolution multiphase model. 
The only difference between the three cases is in how the gas with $n_{\rm H}>0.13$ cm$^{-3}$, the density at which the multiphase model becomes active if on, is treated. 
Since the multiphase model was calibrated to match the observed Kennicutt-Schmidt relation \citep[][]{2003MNRAS.339..289S}, this threshold density corresponds approximately to the density above which stars should start forming \citep[although the exact value depends on some physical assumptions and may depend on redshift; e.g.,][]{2004ApJ...609..667S, 2006ApJ...652..981W}.\\ \\
In principle, the second approach (prescription 7) might seem more accurate since it captures the entire cooling. 
However, as is apparent in the radiative transfer results of \S \ref{lya radiative transfer}, the extra cooling luminosity is concentrated around the accreting galaxy. 
It is unclear whether this central cooling emission would be observable in reality for at least two reasons. 
First, the density of the medium and the immediate proximity of the galaxy imply that locally produced \Lya~photons could be efficiently destructed by dust \citep[the escape fraction of \Lya~photons from Lyman break galaxies (LBGs), for example, covers the entire range $\sim10^{-3}-1$; e.g.,][]{2010ApJ...711..693K}. 
In itself, this is not necessarily an issue for this study in which we focus on a simplified dust-free problem, and would be a well posed problem for follow up studies in which dust would be included. 
Since the extra cooling luminosity occurs in ISM gas, it should however be accompanied by stellar emission which would most likely swamp it locally (see \S \ref{discussion}), and in that case is not really cooling luminosity from the cold streams. 
Second, turning off the multiphase model removes ISM pressurization, which can lead to catastrophic collapse of the gas rich discs and deepen the potential wells, allowing extra energy release. 
Gravitational interactions with dark matter clumps might also artificially transfer energy to unstable gas discs. 
Since this prescription assumes that all the baryons are in the gaseous component, while in the central galaxies of actual galaxies in massive halos a large fraction would be locked in stars, it is likely an upper limit. Prescriptions 8 and 9 are more conservative as they exclude all cooling emission occurring at densities $n_{\rm H}>0.13$ cm$^{-3}$. 
We expect these three cases to bracket the true cooling luminosity.
A better understanding of the energy release at disc interfaces and within galaxies will likely be required to make more definite predictions and should be addressed in future work.\\ \\
We conclude by noting that some caution is in order when interpreting the quantitative details of the cooling luminosity predictions for the halos at the low end of the mass range in Figure \ref{Lalpha vs M}, since they contain relatively few SPH particles ($\sim200$ for $M_{\rm h}=10^{10}$ M$_{\odot}$) and may not be well converged. 
Note, however, that for the realistic prescriptions 7$-$9, the slope of the numerically predicted $L_{\alpha}-M_{\rm h}$ relation agrees well with the analytic expectation based on energy conservation (dashed lines in the Figure), in this regime where the cold mode dominates and where the scaling should apply. 

\begin{figure*}[ht] 
\begin{center} 
\mbox{
{\includegraphics[width=1.0\textwidth]{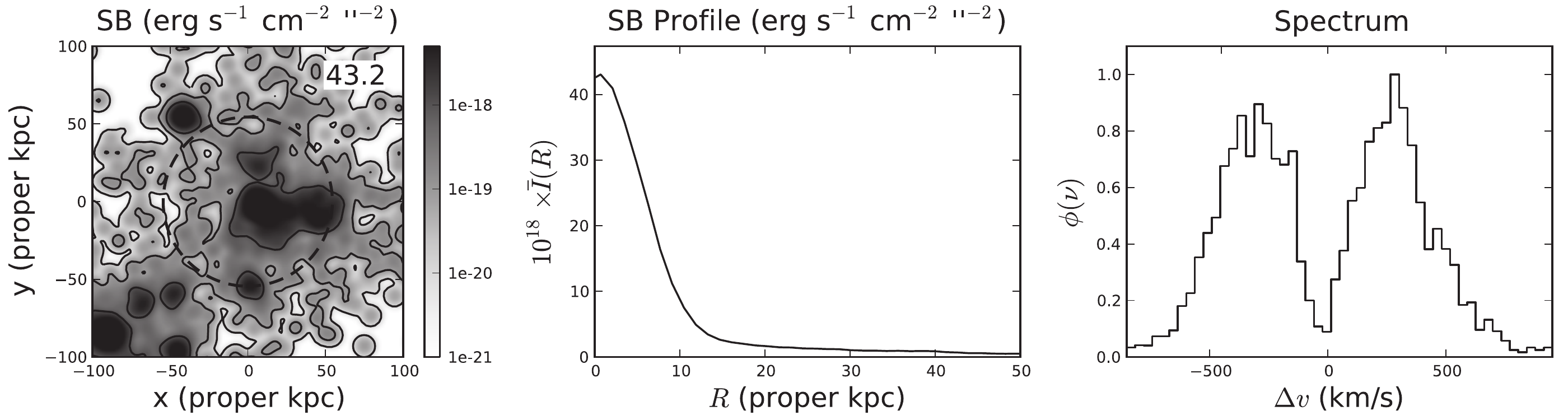}}
}\\
\mbox{
{\includegraphics[width=1.0\textwidth]{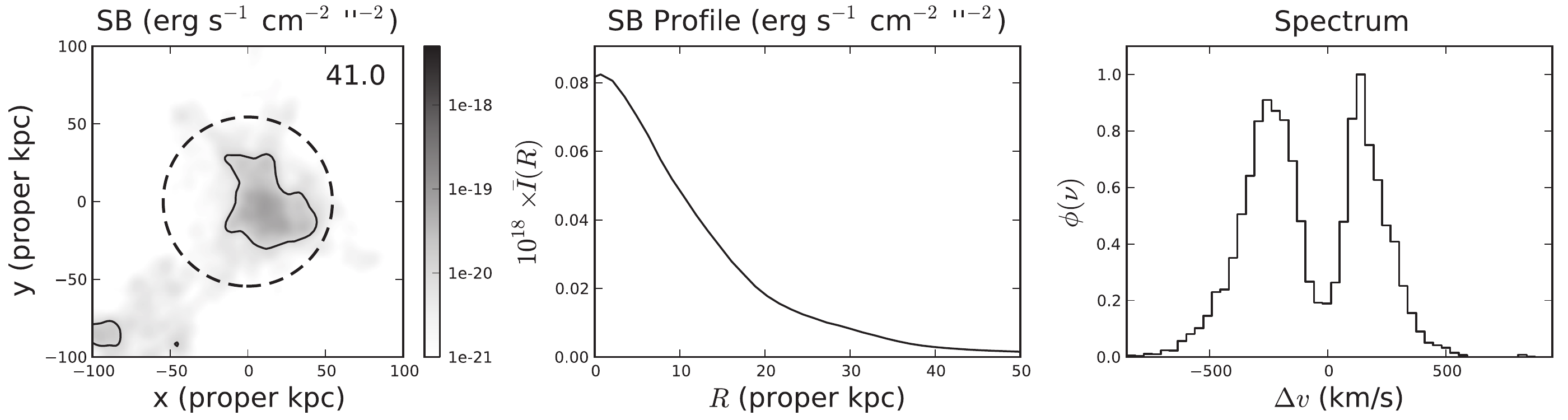}}
}\\
\mbox{
{\includegraphics[width=1.0\textwidth]{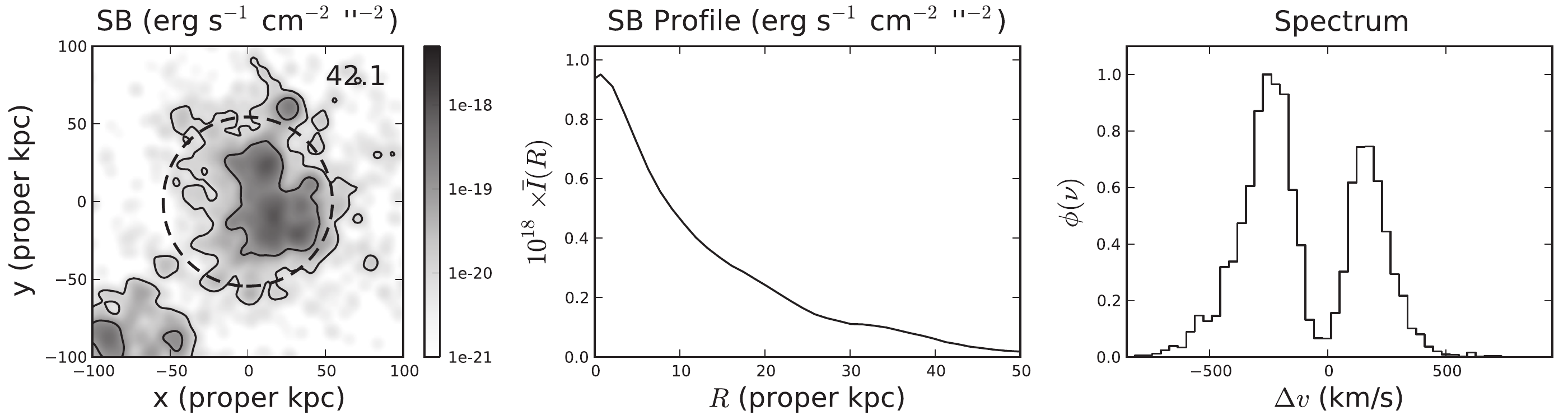}}
}\\
\end{center} 
\caption{Dependence of the \Lya~properties of the A1 system at $z=3$ for prescriptions 1, 3, and 4 for the state of the dense gas. 
The left column shows the observed surface brightness distribution (smoothed with a Gaussian of $FWHM=1''$), the middle column shows the corresponding circularly averaged surface brightness profile, and the right column shows the line spectrum integrated within the virial radius of the halo, indicated by the dashed circles. 
The surface brightness contours correspond to 10$^{-18}$, 10$^{-19}$, and 10$^{-20}$ erg s$^{-1}$ cm$^{-2}$ arcsec$^{-2}$. 
The logarithm of the total apparent line luminosity (in erg s$^{-1}$) within the virial radius is indicated at the top right corner of each surface brightness panel. 
\emph{Top:} (prescr. 1) Standard hydrodynamical simulation, with uniform ionizing background and a multiphase model for star formation, naively integrated over all the particles.  
\emph{Middle:} (prescr. 3) Same simulation, but post-processed with ionizing radiative transfer (\S \ref{post processing self shielding}). The self-shielded regions are regions assumed to be in CIE with the temperature set to $T_{\rm CIE}=10,000$ K.
\emph{Bottom:} (prescr. 4) Same but with self-shielded regions assumed to be in CIE with the temperature set to $T_{\rm CIE}=15,000$ K.}
\label{A1 RT vs physics standard} 
\end{figure*}

\begin{figure*}[ht] 
\begin{center} 
\mbox{
{\includegraphics[width=1.0\textwidth]{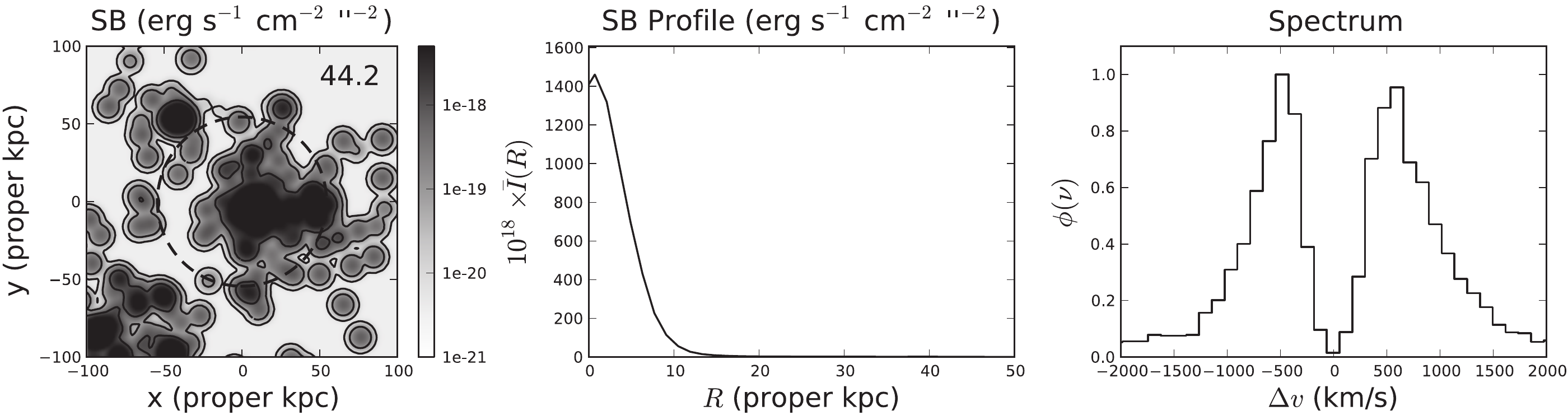}}
}\\
\mbox{
{\includegraphics[width=1.0\textwidth]{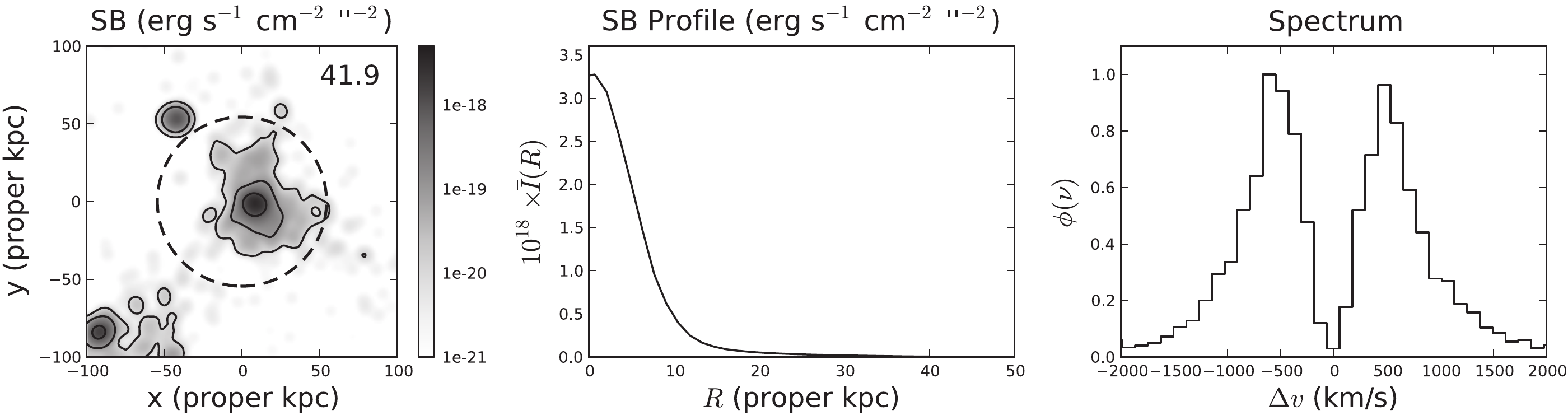}}
}\\
\mbox{
{\includegraphics[width=1.0\textwidth]{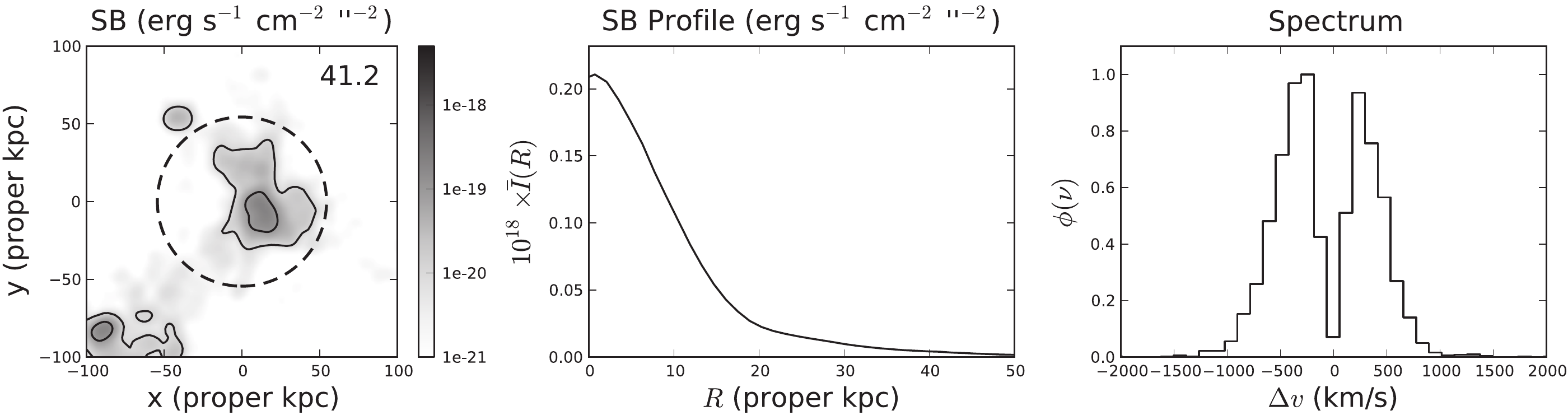}}
}\\
\end{center} 
\caption{Same as Figure \ref{A1 RT vs physics standard}, but for prescriptions 5, 7, and 9. 
\emph{Top:} (prescr. 5) Simulation with an uniform ionizing background but no star formation, integrated over all the particles. 
\emph{Middle:} (prescr. 7) Simulation with the on-the-fly self-shielding approximation and no star formation, integrated over all the particles. 
\emph{Bottom:} (prescr. 9) Simulation with the on-the-fly self-shielding approximation and star formation, but with no \Lya~luminosity from star-forming particles. Note the different velocity scale in comparison with Figure \ref{A1 RT vs physics standard}}
\label{A1 RT vs physics conditional} 
\end{figure*}

\begin{figure*}[ht] 
\begin{center} 
\mbox{
{\includegraphics[width=1.0\textwidth]{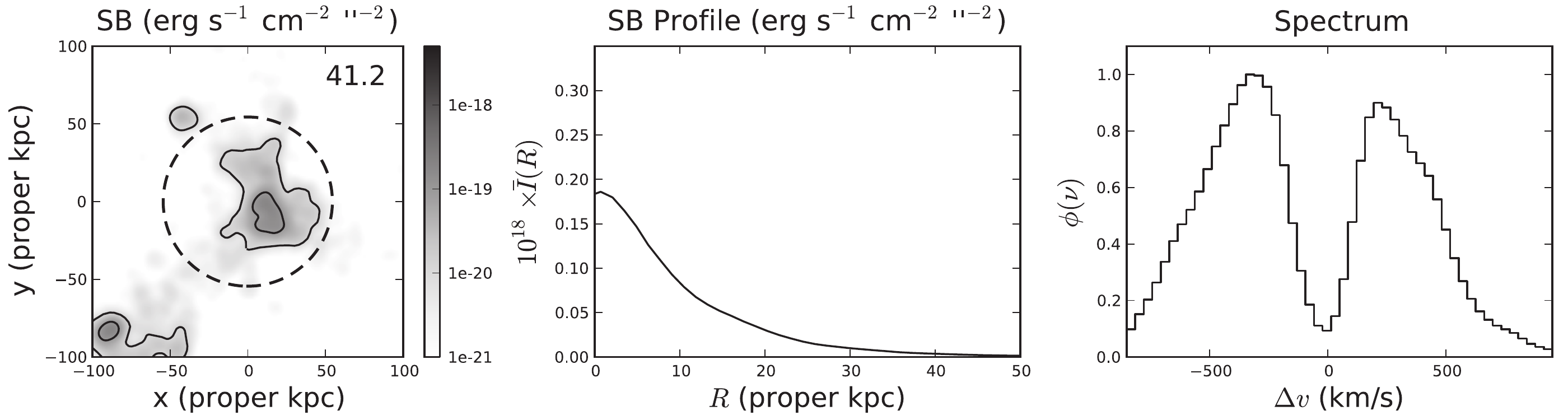}}
}\\
\mbox{
{\includegraphics[width=1.0\textwidth]{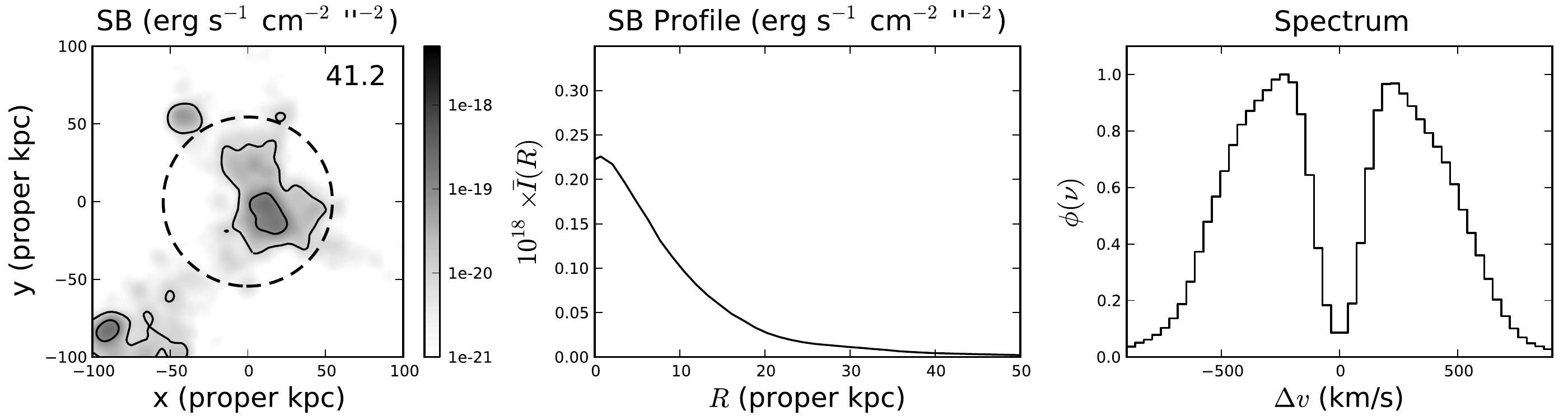}}
}\\
\mbox{
{\includegraphics[width=1.0\textwidth]{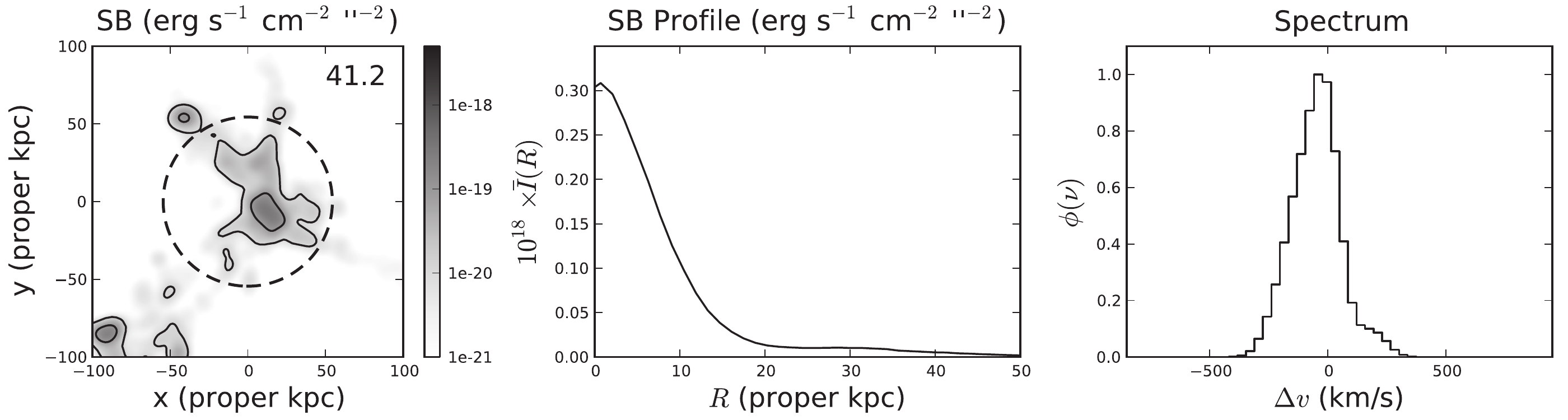}}
}\\
\end{center} 
\caption{Illustration of the effects of radiative transfer on the morphology and spectrum of the \Lya~cooling radiation, for prescription number 9 (on-the-fly self-shielding approximation with the multiphase star formation model, but with the multiphase particles excluded from the luminosity calculation) on the A1 system at $z=3$.
\emph{Top:} Fiducial calculation, with all radiative transfer effects (repeated from Fig. \ref{A1 RT vs physics conditional} but with different axis scales).  
\emph{Middle:} Bulk velocities artificially set zero. The double peaked spectrum is now symmetric since the velocity flows no longer break the symmetry between the blue peak and the red peak. The effective line width is set by the physics of self-shielding (which determines the HI column densities) and of \Lya~radiative transfer. In particular, it has little to do with the global properties of the host halo.
\emph{Bottom:} Bulk velocities back on, but resonant scatters artificially turned off. The morphology on the sky is now sharper since there is no longer spatial diffusion of the photons. More importantly, the spectrum has lost its double peaked nature and its width is set by completely different physics, being in this case representative of the velocity dispersion of the emitting gas.}
\label{A1 RT vs RT physics conditional SF excluded pp} 
\end{figure*}

\begin{figure*}[ht] 
\begin{center} 
\mbox{
{\includegraphics[width=1.0\textwidth]{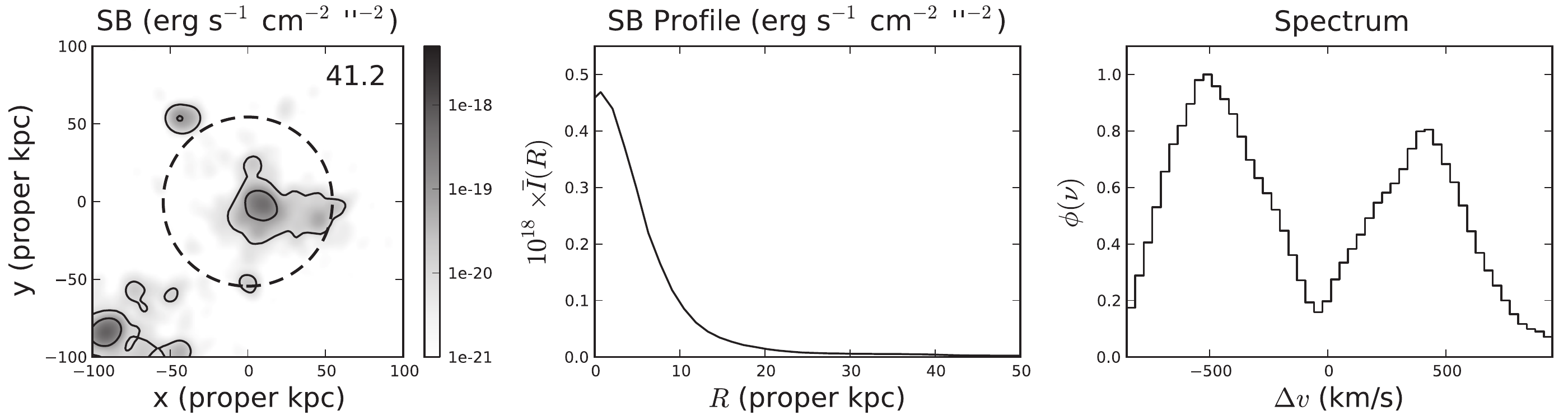}}
}\\
\mbox{
{\includegraphics[width=1.0\textwidth]{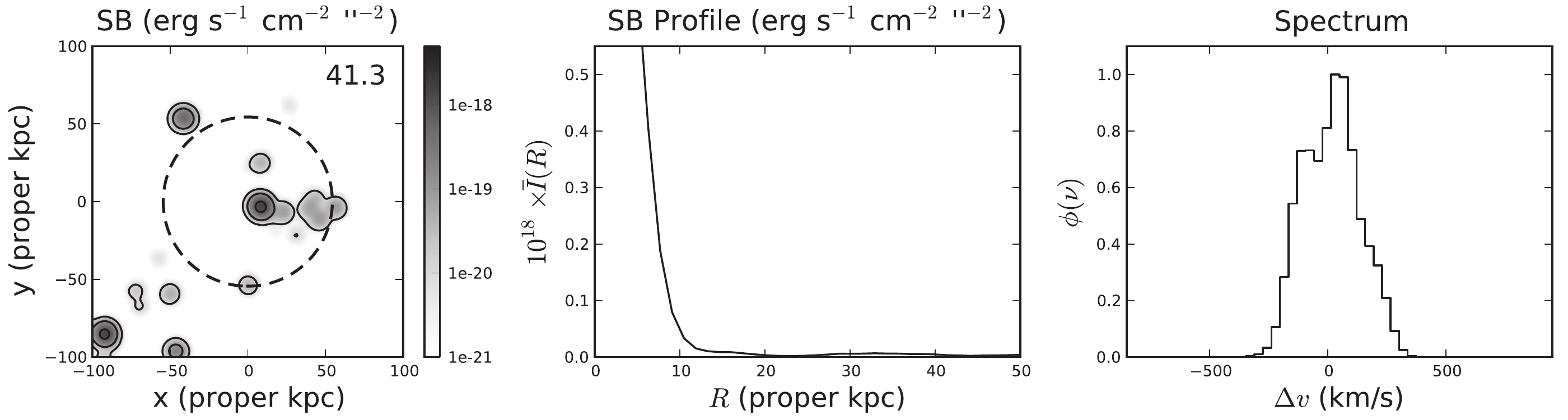}}
}\\
\end{center} 
\caption{Cleaner illustration of the spatial diffusion effect owing to the \Lya~resonant scatters. 
\emph{Top:} Same physical set up as for the cooling radiation prescription number 9 (Fig. \ref{A1 RT vs RT physics conditional SF excluded pp}) but with all the \Lya~photons emitted at the locations of the star-forming particles, in number proportional to the star formation rate. 
To facilitate the visual comparison of the effect on morphology, the total \Lya~luminosity has been normalized to the cooling luminosity for prescription 9.  
\emph{Bottom:} Same calculation but with resonant scattering turned off to show the intrinsic compactness of the star-forming sources.
}
\label{A1 RT vs RT physics conditional SF excluded pp SF only} 
\end{figure*}

\section{Ly$\alpha$ RADIATIVE TRANSFER}
\label{lya radiative transfer}
Having described our hydrodynamical simulations and the modeling of \Lya~photon production, we proceed to the \Lya~radiative transport problem. 
We use a new three-dimensional \Lya~radiative transfer code, $\alpha RT$, described in more detail in Appendix \ref{rt code}.
To summarize, the fields defining the physical state of the gas from a simulation are interpolated onto a Cartesian grid placed around a halo of interest, as for the post-processing ionizing radiative transfer (\S \ref{self shielding}). 
Monte Carlo \Lya~photons are then seeded throughout the gridded volume, with a number proportional to the local \Lya~emissivity (\S \ref{lya emission}).
The multiple resonant scatters of each Monte Carlo photon are simulated until escape.
As the photons propagate, the \Lya~image and corresponding spectrum in each pixel on the sky are constructed as seen by an observer on Earth, taking into account cosmological surface brightness dimming. 
We do not however model the effects of IGM filtering in this work (see \S \ref{radiative transfer physics}), so that the results are more properly interpreted as the redshifted emission as it emerges from the galactic halos.
\\ \\
In typical astrophysical situations, the optical depth at the center of the \Lya~resonance is very large, easily $\tau_{0}>10^{6}$ or more (eq. \ref{lya tau0}).
As a result, a photon propagates only a short distance before being absorbed by and exciting another neutral hydrogen atom to its $2p$ state.
Because of the high Einstein A coefficient for the $2p\to1s$ transition, another \Lya~photon is quickly reemitted ($A_{21}^{-1}\sim10^{-9}$ s).
Since the reemission is in general in a different direction than the incident photon, the propagation can be effectively pictured as a single photon being scattered and undergoing a random walk.
Even if the scattering is coherent in the frame of the scattering atom, the motion of the atom combined with the redirection of the photon in general results in a small shift in the frequency as viewed by an external observer.
This results in a random walk in frequency space as well.
The frequency-space random walk plays a crucial role in shaping the emergent line profile as a photon tends to escape the medium when it finds itself sufficiently far from the line center that the optical depth it sees is reduced to $\sim1$ (unless the medium is so optically thick that the photon spatially diffuses out first).
The transport of \Lya~radiation is thus very different than ordinary lines and a simplified optically thin treatment would lead to fundamental errors both in the theoretical predictions and in interpreting observations.\\ \\
In this section, we present our basic results on \Lya~cooling emission radiative transfer. 
As the focus of this work is to understand the theoretical and numerical uncertainties, and the relevant physical effects, we limit ourselves to examining a particular example, our A1 halo at $z=3$. 

\subsection{Emission Physics}
In \S \ref{lya emission}, we demonstrated how the predicted \Lya~cooling luminosity depends on the treatment of self-shielding and of sub-resolution physics. 
We illustrate how the different possible assumptions manifest themselves in other observational properties by performing \Lya~radiative transfer calculations for some of the prescriptions studied above (Table \ref{prescriptions}). 
The results are shown in Figures \ref{A1 RT vs physics standard} (prescriptions 1, 3, and 4) and \ref{A1 RT vs physics conditional} (prescriptions 5, 7, and 8). 
These fiducial calculations use $N_{\rm ph}=10,000$ Monte Carlo photons. 
For the radiative transfer plots, we subtracted the peculiar motion of the central galaxy with respect to the simulation box (40 km s$^{-1}$ toward the observer), so that $\Delta v=0$ corresponds to the galaxy rest frame.  
As we had found from our study of a sample of halos in \S \ref{self shielding}, incorrectly treating self-shielding (either by assuming an uniform ionizing background, or post-processing ionizing radiative transfer with default simulation temperatures) or including gas particles contaminated by the effective sub-resolution multiphase model can lead to order-of-magnitude overestimates of the cooling luminosity. 
However, it is not only the total luminosity of a system that can be incorrectly predicted, but also its morphology and line spectrum.\\ \\
It is easy to understand how different prescriptions for seeding the \Lya~photons can affect the observed morphology and spectrum. 
In terms of morphology, different prescriptions seed the photons not only in different amounts, but also in different places. 
For instance, including artificially luminous multiphase gas particles produces a disproportionally large \Lya~luminosity concentrated in the densest central parts of the system, approximating a bright point source rather than spatially extended emission from the cold streams.
Seeding the photons in different places also has implications for the emergent spectrum: photons seeded deep within dense self-shielded regions have a much larger HI column density to traverse before escape, and so result in more widely separated double peak profiles.\\ \\
Some specific points are worth noting. 
As indicated by the surface brightness maps and the circularly averaged surface brightness profiles, the resulting objects are spatially extended by tens of proper kpc, comparable to the virial radius of the halo \citep[see also][]{2009MNRAS.400.1109D} and to many of the observed \Lya~blobs \citep[e.g.,][]{2004AJ....128..569M, 2008ApJ...675.1076S}. 
This spatial extent results from a combination of some of the cooling radiation being emitted in the accretion streams far from the galaxy, and of spatial diffusion owing to resonant scattering.  
However, the spatial extent is strongly dependent on the surface brightness threshold and consequently on the luminosity prescription, so that even an intrinsically diffuse source could appear relatively compact in observations
For instance, for our optimistic prescription 7 for the A1 halo at $z=3$ shown in the second row of Figure \ref{A1 RT vs physics conditional}, only the central few kpc would stick out above the surface brightness threshold of $\approx 2\times10^{18}$ erg s$^{-1}$ cm$^{-2}$ arcsec$^{-2}$ of the narrowband images of \cite{2004AJ....128..569M}. 
In this case, the more diffuse cooling halo would be completely missed, but would show up over a larger area in deep long slit spectra sensitive to lower surface brightnesses \citep[e.g.,][]{2008ApJ...681..856R}. 
The predicted line spectra are double peaked, a common characteristic of \Lya~radiative transfer reflecting the fact that the photons can escape the medium either on the blue side or on the red side of the line center, where the opacity is typically too extreme \citep[e.g.,][]{1990ApJ...350..216N, 2002ApJ...578...33Z, 2006ApJ...649...14D, 2006A&A...460..397V}. 
In most cases (but not all, reflecting the effects of the complex geometry in the different prescriptions), the blue peak is slightly more pronounced than the red peak. 
This is a signature of systematic infall in the problem at hand, in which the velocity gradients tend to smear the line opacity on the red side of the \Lya~line, making it easier for the photons to escape on the blue side.
Almost all the star-formation powered \Lya~emitters \citep[][]{2003ApJ...588...65S, 2010arXiv1003.0679S}, as well as many \Lya~blobs and fainter analogues \citep[][]{2006ApJ...640L.123M, 2008ApJ...675.1076S, 2008ApJ...681..856R}, instead show dominant red peaks indicative of outflows. 
We do not see this phenomenon here simply because we have not modeled galactic winds in our simulations to simplify the physical problem. 
Unlike the thin accretion streams (with small covering factor) that produce only slightly stronger blue peaks, outflows (with order unity covering factor) are expected to boost the red peak more drastically \citep[e.g.,][]{2006ApJ...649...14D, 2006A&A...460..397V}. 
Intergalactic absorption, neglected here, would also tend to preferentially suppress the blue peak. 
In future work, we will include outflows and IGM filtering, which should provide a better match to the observational data.\\ \\
In \S \ref{lya emission}, we had identified three physically plausible prescriptions for calculating the \Lya~cooling luminosity (7, 8, and 9); the radiative transfer results for 7 and 9 are shown in the second and third rows of Figure \ref{A1 RT vs physics conditional}. 
The surface brightness maps now make it clear that the extra luminosity obtained using prescription 7, i.e. when using the on-the-fly self-shielding approximation but turning off the star formation model and summing over the all the particles, is concentrated in the densest central parts of the system. 
This also manifests itself in the predicted line spectrum, which shows more widely separated peaks as a result of the higher column densities of the gas through which the bulk of the photons must propagate to escape, as discussed in the following section. 

\subsection{Radiative Transfer Physics}
\label{radiative transfer physics}
Since radiative transfer effects play a crucial role in shaping the observational properties, we pause to illustrate exactly how important each piece of physics is. 
To do so, we repeat the same fiducial calculations but sequentially turn off different physical effects. 
We only repeat the calculations for the plausible prescription 9, which suffices to illustrate the physics. 
The results are presented in Figure \ref{A1 RT vs RT physics conditional SF excluded pp}.\\ \\
First, we keep resonant scatters but artificially set the gas velocities to zero. 
In this case, the most important change is in the line spectrum, which becomes a nearly symmetric double peak. 
This is easily understood in the context of the plane-parallel analytic solution derived by \cite{1990ApJ...350..216N} \citep[for an adaptation to spherical geometry, see][]{2006ApJ...649...14D} for a monochromatic source in an extremely optically thick, static medium. 
Assuming that the source is located at the center of the slab and that $\tau_{0}$ is the line center optical depth from the source to the surface, \cite{1990ApJ...350..216N} showed that the emergent spectrum is a symmetric double peak profile, with each peak offset by
\begin{equation}
\label{Delta v Neufeld}
|\Delta v_{\rm p}| \approx 191~{\rm km~s^{-1}} 
\left( \frac{T}{10^{4}~{\rm K}} \right)^{1/6}
\left( \frac{N_{\rm HI}}{10^{20}~{\rm cm^{-2}}} \right)^{1/3}
\end{equation}
from the center. 
In dimensionless units in which $x\equiv(\nu - \nu_{0})/\Delta \nu_{D}$ is the frequency offset in Doppler-broadening units, $|x_{\rm p}| = 1.06 (a \tau_{0})^{1/3}$. 
For convenience, the corresponding offset in observed wavelength units can be obtained from 
\begin{equation}
\Delta \lambda_{\rm obs} \approx 8.1~{\rm \AA}
\left( 
\frac{1+z}{4}
\right)
\left(
\frac{\Delta v}{500~{\rm km~s}^{-1}}
\right).
\end{equation}
In particular, the emergent line profile and corresponding effective line width are in this case determined by the physics of \Lya~radiative transfer and of self-shielding (which sets the HI column densities) and have little to do with the global properties of the host halo.\\ \\
Next, we turn velocities back on but artificially turn off resonant scattering. 
This is equivalent to assuming that the \Lya~line were an ordinary, optically thin line. 
In this case, there are two important changes. 
First, the morphology on the sky is slightly sharper since there is no longer a spatial diffusion effect arising from photon random walks. 
The spatial diffusion effect appears subtle here as the cooling emission is intrinsically diffuse. 
It is better illustrated in Figure \ref{A1 RT vs RT physics conditional SF excluded pp SF only}, in which we have kept the same physical set up but seeded the \Lya~photons in proportion to the local star formation rate, which produces a much more intrinsically compact source. 
To facilitate visual comparison of the morphology with the pure cooling cases, we have normalized the star formation \Lya~emission to the cooling luminosity of the halo. 
This halo is however forming stars at a rate $\sim30$ M$_{\odot}$/yr, which could result in up to $\sim3\times10^{43}$ erg s$^{-1}$ of star formation powered \Lya~emission if the escape fraction were unity \citep[][]{1999ApJS..123....3L}; in this case, it would dominate over the cooling luminosity of the halo. 
Second, the line spectrum is very different and has essentially lost its double peaked nature. 
Most importantly, the spectrum (particularly its width) is determined by completely different physics, since it is now directly representative of the velocity dispersion within the host halo through the Doppler effect.\\ \\
An important implication of the radiative transfer effects on the observed \Lya~line width concerns the way the masses of extended \Lya~sources like the \Lya~blobs are estimated. 
In fact, it is often assumed that the \Lya~line width can be associated with random motion within the halo \citep[e.g.,][]{2004MNRAS.351...63B, 2006ApJ...640L.123M}. 
As could be anticipated from more general \Lya~radiative transfer studies \citep[e.g.,][]{2002ApJ...578...33Z, 2006ApJ...649...14D, 2006A&A...460..397V}, our results show that neglecting radiative transfer effects will tend to overestimate the mass, since they alone broaden the \Lya~line even for completely static emitting media. 
However, the situation is in reality more complex since the \Lya~lines that one actually observe on Earth are further filtered by the intervening IGM, which can significantly attenuate the \Lya~line \cite[e.g.,][]{2010ApJ...716..574Z, 2010arXiv1003.4990Z, 2010arXiv1009.1384L}. 
At $z=3$, for example, the diffuse IGM transmits only $\approx67$\% of photons immediately blueward of \Lya~and this fraction decreases rapidly with increasing redshift \citep[e.g.,][]{2008ApJ...681..831F}. 
Moreover, local matter overdensities and systematic infall around massive halos \citep[e.g.,][]{2008ApJ...673...39F} can amplify and shift the absorption, whereas galactic winds act to ``let through'' more radiation by redshifting it away from the \Lya~line center \citep[e.g.,][]{2004MNRAS.349.1137S, 2007MNRAS.377.1175D, 2010arXiv1004.2490D}. 
It is therefore in general difficult to relate the observed \Lya~line alone to the properties of the halo producing it, and it is not even clear in any given case whether the observed line width over$-$ or underestimates the halo velocity dispersion.  
More systematic studies quantifying the relation between the observed \Lya~line properties and the host halo mass, extending the type of calculations presented in this work, are certainly warranted. 
We here simply caution against taking halo masses estimated from the \Lya~line too literally.
This remark applies even if the \Lya~blobs are not predominantly powered by cooling radiation since radiative transfer effects will necessarily be at play in any \Lya~source.

\begin{figure*}[ht] 
\begin{center} 
\mbox{
{\includegraphics[width=1.0\textwidth]{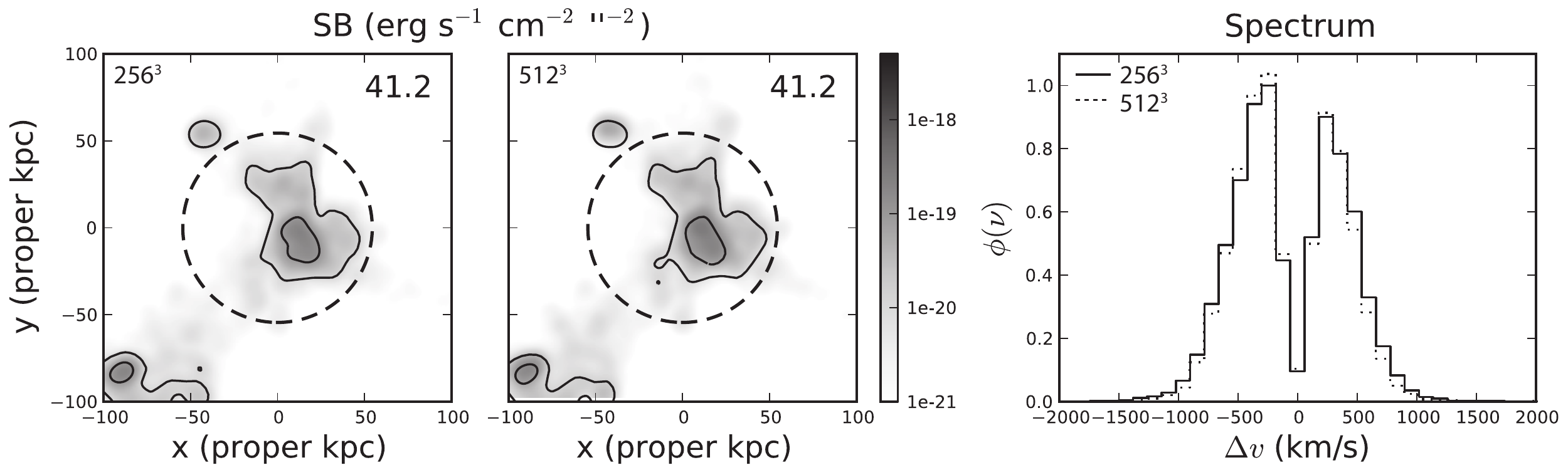}}
}\\
\end{center} 
\caption{Illustration of the convergence of our radiative transfer results with the number of grid points along each dimension. 
\emph{Left:} Fiducial cooling luminosity calculation for prescription number 9, with $256^{3}$ radiative transfer grid points. 
\emph{Middle:} Same, but with $512^{3}$ radiative transfer grid points in the same volume. 
\emph{Right:} Corresponding line spectra within the virial radius.
Both the morphology and the line spectrum appear well converged with the radiative transfer resolution.}
\label{convergence fig} 
\end{figure*}

\subsection{Convergence}
\label{convergence section}
Most of our radiative transfer calculations employed physical fields stored on a Cartesian grid with $256^{3}$ cells in a volume of (1 comoving Mpc/h)$^{3}$. 
Since the \Lya~photon trajectories are not tied to the grid points (see Appendix \ref{rt code}), their increments were however finer than that this by a factor of 5. 
Furthermore, the cell luminosities were always calculated directly from the SPH particles so that they captured the full gas clumping and were not subject to resolution degradation. 
To verify that our radiative calculations are robust, we have repeated many of them with $512^{3}$ grid points in the same volume instead. 
As shown in Figure \ref{convergence fig} for prescription 9 applied to the A1 halo at $z=3$, the calculations in fact appear well converged both in terms of \Lya~morphology and spectrum.

\section{DISCUSSION AND CONCLUSION}
\label{discussion}
Motivated by observations of a variety of extended \Lya~sources at high redshift and by recent studies suggesting that cold mode accretion could account for the majority of them \citep[e.g.,][]{2009MNRAS.400.1109D}, we have used hydrodynamical simulations to predict the \Lya~cooling emission of forming galaxies. 
We have introduced a new \Lya~radiative transfer code, $\alpha RT$, and for the first time applied such a code to the particular problem of cooling radiation in cosmological hydrodynamical simulations.
In this work, which will serve as the foundation for follow up studies of the astrophysical implications, we have quantified the theoretical and numerical uncertainties in predicting the properties of \Lya~cooling radiation. 
To do so, we have considered the simplest physical problem of galaxy assembly embedded in the cosmic UV background but without feedback processes. 
We have shown that the \Lya~cooling luminosities, morphologies, and spectra of the cold streams are strongly dependent on the treatment of the ionization and thermal state of the emitting gas, and that naive assumptions can produce artificially high cooling luminosities. 
Previous studies have usually relied on such assumptions, because most existing hydrodynamical simulations do not self-consistently follow the transfer of ionizing radiation \citep[but see][]{2006ApJ...644L...1S, 2007ApJ...657L..69L, 2009ApJ...704.1640L}, whereas the cold streams in reality self-shield.
The predicted cooling luminosity is so sensitive to the thermal state of the gas principally because of the exponential dependence of the collisional excitation coefficient on temperature in the range $T\sim10^{4}-10^{5}$ K characteristic of the cold streams. 
We have also demonstrated that subresolution physics models, when not carefully taken into account, can lead to large errors as well.\\ \\
Having explicitly illustrated the range of uncertainties, we have made a systematic attempt to converge on the correct answer. 
Our strategy consisted of first post-processing our hydrodynamical simulations with ionizing radiative transfer to identify the self-shielded regions. 
By inspection, we found that there is a fairly well defined total hydrogen density $n_{\rm H}\sim0.01$ cm$^{-3}$ above which the gas self-shields, at least at the redshifts $z \sim 3$ characteristic of the existing observations. 
We have then rerun simulations with exactly the same initial conditions and other physical parameters, but with the ionizing background turned off in regions exceeding the fiducial density threshold $n_{\rm H}=0.01$ cm$^{-3}$. 
By turning the ionizing background off in dense regions on the fly, this dense gas is consistently evolved with modified heating and cooling functions, as well as the corresponding dynamical response. 
These simulations provide us with our best estimates for the actual \Lya~cooling luminosity of the cold streams, in good agreement with energetic considerations.\\ \\
For our \Lya~radiative transfer calculations, we focused on a particular halo (A1) of total mass $M_{h}=2.5\times10^{11}$ M$_{\odot}$ at $z=3$. 
We have shown that it is not only the integrated \Lya~luminosity of the system that depends strongly on assumptions regarding the thermal state of the gas, but also its apparent morphology and spectrum. To isolate the role of different physical effects in shaping the observational properties, we have sequentially turned off separate pieces of physics. 
Both the \Lya~resonant scatters as well as the bulk velocity structure of the system are critical in shaping the emergent spectrum. 
The resultant effective line width in general differs strongly from the optically thin expectation owing to these radiative transfer effects, and we therefore caution against measuring the masses of systems based on the \Lya~line alone. 
The intrinsically extended nature of the cooling emission and the spatial diffusion owing to resonant scattering combine to produce objects that are spatially extended on the sky, but with an apparent size that depends sensitively on the observational surface brightness threshold. 
As a result, only the central few kpc of the A1 halo \Lya~cooling emission at $z=3$ sticks out above the surface brightness threshold $\sim10^{-18}$ erg s$^{-1}$ cm$^{-2}$ arcsec$^{-2}$ typically achieved to date (Fig. \ref{A1 RT vs physics conditional}). 
If the characteristic size of the \Lya~emission scales as $R\propto M_{\rm h}^{1/3}$, then the surface brightness should scale as $SB \propto L_{\alpha} / R^{2} \propto L_{\alpha} / M_{\rm h}^{2/3}$. 
For our consistent prescriptions 7$-$9, Figure \ref{Lalpha vs M} shows that the high-mass luminosity slope varies from about 3/4 to 1 in the most optimistic case, i.e. $SB \propto M_{\rm h}^{-0.33...-0.08}$, so that even the higher-mass systems are unlikely to show up as sources with extents $\sim100$ kpc in existing observations from cooling luminosity alone. 
Fainter sources may however well be detectable in deeper observations, such as the 100-hr long-slit spectrum reported by \cite{2008ApJ...681..856R}.\\ \\
Our analysis of a sample of halos from a cosmological volume at $z\approx3$ confirms that \citep[in agreement with previous studies;][]{2000ApJ...537L...5H, 2001ApJ...562..605F, 2009MNRAS.400.1109D, 2010MNRAS.407..613G} cooling emission alone can \emph{in principle} produce luminosities $L_{\alpha} \sim 10^{43}-10^{44}$ erg s$^{-1}$ sufficient to account for luminous \Lya~blobs. 
This however requires quite optimistic assumptions under which most of the energy is released close to, or within, the accreting discs, at densities sufficient to form stars according to the observed Kennicutt-Schmidt relation.
In those optimistic predictions based on prescription 7, star formation was artificially turned off (yielding purely gaseous discs), so that it is unlikely that this entire cooling emission can be realized in reality without being overwhelmed by stellar emission.\footnote{In their simulations of LBGs at $z=3.6$, Laursen et al. 2009a find that only $\sim10$\% of the total \Lya~emission comes from cooling, although we show in Appendix \ref{analytic considerations appendix} that this ratio should depend on halo mass and redshift.}
Our findings in this respect are at odds with the recent simulations of \cite{2010MNRAS.407..613G}, who argue that pure cooling radiation can explain the observed giant \Lya~blobs. 
Our investigation of the sources of error in the \Lya~luminosity predicted from simulations (\S \ref{lya emission}) allows us to understand the differences. 
First, we have demonstrated that the predicted \Lya~luminosity is very sensitive to the treatment of self-shielded gas and that at $z\approx3$, the characteristic density above which gas self-shields from the UVB is $n_{\rm H}\approx0.01$ cm$^{-3}$. \cite{2010MNRAS.407..613G} did not explicitly perform ionizing radiative transfer and instead assumed a higher self-shielding threshold of $n_{\rm H}=0.1$ cm$^{-3}$. 
This higher density threshold translates into an overestimate of the \Lya~luminosity owing to artificial photoheating of gas in the density range $0.01\leq n_{\rm H} \leq 0.1$ cm$^{-3}$. 
Examination of the luminosity-weighted $T-n_{\rm H}$ histogram in their Figure 7 in fact indicates that the bulk of the \Lya~luminosity they predict originates from this density regime (including in clumps that would correspond to satellite galaxies, even when the central galaxy is excised), whereas \Lya~emission is strongly suppressed at these densities in our approximation to the consistent thermal evolution of self-shielded gas, owing to rapid cooling below $10^{4}$ K (Fig. \ref{T vs nH}). 
While there is some uncertainty in the precise value of the self-shielding threshold, our radiative transfer calculations show that it is clearly below 0.1 cm$^{-3}$ at the redshifts under consideration (Fig. \ref{xHI vs nH}), and the value $n_{\rm H}\approx0.01$ cm$^{-3}$ is further supported by recent radiative transfer calculations by other groups \citep[][]{2010arXiv1006.5345N, 2010arXiv1004.2503A}. 
The cleanest contrast between our results and those of \cite{2010MNRAS.407..613G} is provided by their predictions of the \Lya~luminosity originating from the ``streams'' alone, in which they excluded all emission from within 20\% of the virial radius. 
Those can be compared, at best, with our pessimistic prescriptions 8 and 9, in which only gas with $n_{\rm H}>0.13$ cm$^{-3}$ (above the subresolution star formation threshold) is excluded; our predictions are in those cases a factor $\gtrsim10$ lower than theirs. 
It would not be fair to compare our most optimistic prescription 7 with the \cite{2010MNRAS.407..613G} results with the central 20\% of the virial radius excluded, since most of the extra emission under this prescription occurs very close to, or inside, the central galaxy. 
Finally, the simulations of \cite{2010MNRAS.407..613G} included supernova feedback, which provides an additional source of energy and could also contribute significantly to boosting the cooling luminosity, even in the absence of ionizing photons from associated local sources  \citep[e.g.,][]{2000ApJ...532L..13T, 2001ApJ...562L..15T, 2003ApJ...591L...9O, 2004ApJ...613L..97M}.\\ \\
It is further notable that many \Lya~blobs show spectral signatures of outflows (dominant red peak), at odds with the expectation of a more prominent blue peak in the case of pure cooling of infalling gas.\footnote{There are exceptions, for example certain regions of the LAB2 blob at $z=3.09$ \citep[][]{2005Natur.436..227W}.} 
Nonetheless, the release of gravitational potential energy through cooling radiation at observationally interesting levels is an inevitable prediction of galaxy formation in $\Lambda$CDM and the growing body of theoretical work supporting the cold mode scenario implies that a significant fraction would come out in \Lya. 
Such sources are therefore poised to be routinely detected in future surveys, and likely account for at least a subpopulation of the more modest extended \Lya~sources, such as the ones detected by \cite{2008ApJ...681..856R}. 
In fact, at least three of the \cite{2008ApJ...681..856R} sources (\#15, 36, and 37) show spectral signatures suggestive of infall (with a more prominent blue peak). Interestingly, GALEX non-detections indicate that the bright \Lya~blobs are much rarer at $z=0.8$ than at $z \gtrsim 2$ \citep[][]{2009AJ....138..986K}. 
This redshift evolution is reminiscent of the gradual disappearance of the cold mode at low redshift predicted by theoretical studies \citep[e.g.,][]{2003MNRAS.345..349B, 2006MNRAS.368....2D, 2005MNRAS.363....2K, 2009MNRAS.395..160K}. 
It is important to note that the cold mode could play an important role in the existence and properties of the \Lya~blobs even if they are not energetically dominated by cooling emission. 
In fact, the presence of cold neutral gas in galactic halos enhances the conversion stellar or AGN power into \Lya~photons, and scattering off such gas may be necessary to produce the morphological and spectral properties observed.\\ \\
Although we have made important progress in accurately predicting the \Lya~properties of cold accretion, much work remains to be done. 
For instance, we have focused on the simplest physical problem of accretion embedded in the cosmic UV background and neglected feedback processes. 
In the actual Universe, galactic winds are observed to be ubiquitous at high redshift \citep[e.g.,][]{2003ApJ...588...65S, 2010arXiv1003.0679S} and simulations suggest that they are in fact needed to reproduce the observed baryonic mass function of galaxies \citep[e.g.,][]{2009MNRAS.396.2332K, 2010MNRAS.406.2325O}. 
Furthermore, AGN can inject large amounts of energy in the surrounding gas \citep[][]{2005MNRAS.361..776S, 2006ApJS..166....1H, 2009arXiv0909.2872D}, which could radiate even after the nucleus has shut off. 
Such processes will modify the kinematic and thermal properties of the circumgalactic medium, and therefore its \Lya~signatures. 
Moreover, they provide additional sources of energy that could enhance the total emission. 
The ionizing radiation produced by embedded star formation or AGN \citep[e.g.,][]{2010ApJ...708.1048K}, the effects of metals on the cooling function \citep[e.g.,][]{2010MNRAS.403L..16C}, and the destruction of \Lya~photons by dust \citep[as shown by][]{2009ApJ...704.1640L} are also likely to be important to reproduce the observed sources, but have not been modeled here for simplicity. 
Since cold accretion likely fuels star formation in halo centers, at rates similar to the cold gas accretion rates \citep[e.g.,][]{2005MNRAS.363....2K, 2009MNRAS.395..160K, baryona}, gas accretion and star formation are intrinsically linked and their roles in \Lya~emission are difficult to decouple. 
Follow up studies will build upon the technical foundation established in this work and include some of these effects. 
Ultimately, cosmological statistics like the luminosity and correlation functions of the \Lya~sources will be predicted. 
It will also be interesting to consider complementary observational probes of the cold streams, including their \Lya~polarization \citep[e.g.,][]{1999ApJ...520L..79R, 2008MNRAS.386..492D}, their absorption signatures, and other emission lines (including from metals).\\ \\
On a more basic level, our understanding of the physics of the cold mode is incomplete. 
In particular, it is not exactly clear how the cold streams release their gravitational energy. 
At least some simulations indicate that they maintain a roughly constant velocity, rather than accelerate, as they fall into galactic halos \citep[][]{2005MNRAS.363....2K}. 
This implies that the streams would continuously radiate the gravitational work done on them, perhaps by undergoing a series of weak shocks. 
However, these weak shocks have not been explicitly identified in existing simulations. \cite{2003MNRAS.345..349B} instead envision a scenario in which the cold streams free fall into the halos and release little energy before hitting the central disc in a strong shock, which may be partially supported by our results which suggest that a large fraction of the energy could be released near the halo center. 
Infalling cold gas might also release some energy through interactions with halo sub-structure and with the lower density, hot halo gas. These interactions could involve hydrodynamic instabilities below the resolution of current generation cosmological hydrodynamic simulations, both SPH and AMR. 
While we have taken the pragmatic point of view of predicting the observational signatures implied by our current galaxy formation models, it is conceivable that the existing simulations are subject to numerical limitations. 
Work is underway to elucidate some of these hydrodynamical issues using the new shock-capturing, moving mesh code AREPO \citep[][]{2010MNRAS.401..791S}. 
When the capability becomes available, it would also be useful to perform truly self-consistent ionizing radiative transfer on the fly in order to definitively capture the thermal evolution of the gas.

\acknowledgements 
We are grateful to Volker Springel for making the  expanded GADGET code available to us, and to Yuval Birnboim, Juna Kollmeier, Adam Lidz, Avi Loeb, Matt McQuinn, Eliot Quataert, Chuck Steidel, and David Weinberg for useful discussions. 
We also thank the referee, Jesper Sommer-Larsen, for a detailed and constructive review. 
CAFG is supported by a fellowship from the Miller Institute for Basic Research in Science, and received further support from the Harvard Merit Fellowship and FQRNT during the course of this work. 
DK is supported by NASA through Hubble Fellowship grant HST-HF-51276.01-A awarded by the Space Telescope Science Institute, which is operated by the Association of Universities for Research in Astronomy, Inc., under contract NAS 5-26555. 
Partial funding was also provided by NSF grants ACI 96-19019, AST 00-71019, AST 02-06299, AST 03-07690, AST 05-06556, AST 05-06556, AST 09-07969, and PHY 08-55425, and NASA ATP grants NAG5-12140, NAG5-13292, NAG5-13381, and NNG-05GJ40G. 
Further support was provided by the David and Lucile Packard Foundation, the Alfred P. Sloan Foundation, the John D. and Catherine T. MacArthur Foundation, and Harvard University funds. 
The computations in this paper were run on the Odyssey cluster supported by the FAS Sciences Division Research Computing Group at Harvard University. 

\appendix

\section{A. ANALYTIC CONSIDERATIONS}
\label{analytic considerations appendix}
In this section, we outline analytic arguments that motivate the expectation of significant \Lya~emission from galactic gas accretion, at a level comparable to observed \Lya~blobs.
Similar arguments were made by \cite{2000ApJ...537L...5H}, \cite{2009MNRAS.400.1109D}, and \cite{2010MNRAS.407..613G}; their essence is summarized here as a basis to understand our numerical results. We also provide an analytic estimate of the amount of power that could be contributed by photoionization from the cosmic background.\\ \\
We consider gas accreting from the IGM to the bottom of a dark matter halo potential well and ask how much energy is available to be radiated.
Averaged over the accretion time scale, the gravitational power available can be expressed as $\langle \dot{E}_{\rm grav} \rangle\approx f_{\rm eff} \dot{M}_{\rm gas} |\Delta \Phi_{\rm grav}(r_{\rm min})|$, where $\dot{M}_{\rm gas}$ is the gas mass accretion rate, $\Delta \Phi_{\rm grav}(r_{\rm min})$ is the potential difference from the IGM to the radius $r_{\rm min}$ from the bottom of the potential well at which the gas is assumed to settle, and $f_{\rm eff}<1$ is an efficiency factor.
The efficiency factor accounts the fraction of the gravitational energy that remains in bulk kinetic or thermal form.
In addition, only a certain fraction of the energy that is radiated comes out in the \Lya~line.\\ \\
Using the fitting formula derived by \cite{2008MNRAS.383..615N} for the average halo mass accretion rate, we can write (for the $WMAP5$ cosmology)
\begin{equation}
\label{Mdot eq}
\dot{M}_{\rm gas} \approx 210~{\rm M}_{\odot}{\rm~yr^{-1}} 
\left(
\frac{M}{10^{12}~{\rm M_{\odot}}}
\right)^{1.4}
\left(
\frac{1+z}{4} 
\right)^{2.5}
\left( \frac{f_{\rm gas}}{0.165} \right),
\end{equation}
where $f_{\rm gas}$ is the fraction of the accreted mass that is gaseous \citep[see also][]{2009MNRAS.398.1858M, 2010arXiv1001.2304F}.
For a \cite{1997ApJ...490..493N} halo profile, 
\begin{equation}
\label{Delta Phi eq}
|\Delta \Phi_{\rm grav}(r_{\rm min})| \approx 
\frac{G M_{200c}}{r_{\rm min}}
\frac{\ln{[1+(r_{\rm min}/r_{200c})c]}}{\ln(1+c) - c/(1+c)}
\approx \frac{G M_{200c}}{r_{200c}}
\frac{c}{\ln{(1+c) - c/(1+c)}},
\end{equation}
where $M_{200c}$ is the halo mass defined as that exceeding 200 times the critical density, $r_{200c}$ is the corresponding radius, and $c$ is the halo concentration parameter.
In the last step, we have assumed $c r_{\rm min}/r_{200c} \ll 1$ to simplify the expression.
In general, $M_{200c} \sim M$ ($\equiv M_{\rm FoF} \approx~M_{\rm 180b}$; see \S \ref{simulation parameters and halo identification}), but since $r_{180b}$ may exceed $r_{200c}$ by as much of 75\% on cluster scales \citep[][]{2002ApJS..143..241W}, $M_{200c}$ can be lower than $M$ by a factor of as much as 5 at $z\sim0$.
For the present crude estimate, we will ignore the distinction between $M$ and $M_{\rm 200c}$, which is much smaller at the high redshifts of interest.
Combining equations \ref{Mdot eq} and \ref{Delta Phi eq}, we find
\begin{equation}
\label{Egrav max}
\langle \dot{E}_{\rm grav} \rangle\approx 3.8\times10^{43}{\rm~erg~s^{-1}}f_{\rm eff} 
\left( \frac{M}{10^{12}{\rm~M_{\odot}}} \right)^{1.8}
\left(
\frac{1+z}{4}
\right)^{3.5}
\left(
\frac{f_{\rm gas}}{0.165}
\right)
\left(
\frac{
c/[\ln{(1+c)} - c/(1+c)]
}
{5.2}
\right).
\end{equation}
This expression is valid at $z \gtrsim 1$ (where the cosmological constant can be neglected) and fiducially assumes $c=5$, approximately the median concentration of all massive halos at $z\sim2-3$ \citep[e.g.,][]{2003ApJ...597L...9Z}.\\ \\
As an estimate of the sensitivity of this analytic prediction to the shape of the dark matter halos, the calculation can be repeated for a \cite{1990ApJ...356..359H} profile, 
\begin{equation}
\Phi_{\rm grav}(r)  = - \frac{G M_{200c}}{r + a},
\end{equation}
using the relation $a = (r_{200c}/c) \sqrt{2[\ln{(1+c) - c/(1+c)}]}$ \citep[e.g.,][]{2005MNRAS.361..776S}. 
We then obtain the ratio
\begin{equation}
\frac{\langle \dot{E}_{\rm grav} \rangle^{\rm Hernquist}}{\langle \dot{E}_{\rm grav} \rangle^{\rm NFW}}  = 
\sqrt{\frac{\ln{(1+c) - c/(1+c)}}{2}}.
\end{equation}
For $c=3,~5,~{\rm and}~10$, this ratio takes the values $0.56,~0.69,~{\rm and}~0.86$, respectively. 
Since these differences, at the tens of percent level, are subdominant compared to the other sources of uncertainty, we will only plot the prediction for the NFW model in this work. 
\\ \\
In the absence of additional sources heating (see below, where we consider the importance of the ionizing background), the actual \Lya~cooling luminosity emergent from a given halo will in general be somewhat lower than predicted by equation \ref{Egrav max}, since some of the gravitational power will be converted to kinetic and thermal forms, rather than radiated, and the radiated fraction will not come out entirely in \Lya.
For halos dominated by the cold mode, however, the $f_{\rm eff}$ is expected to be significant, perhaps up to $\sim 50$\% if half of the gravitational energy is radiated and most of this radiation comes out in \Lya. 
At higher masses, the efficiency will be suppressed as the hot mode becomes more important and a higher fraction of the baryons are accreted in the form of stars.\\ \\
The analytic model of \cite{2009MNRAS.400.1109D} is more sophisticated than the simple energetic argument above, as it accounts for the fact that the fraction of cold gas in halos diminishes with increasing mass. 
Since only the cold gas radiates efficiently in \Lya, this yields a shallower slope for the $L_{\alpha}-M_{h}$ relation at high masses, which is in fact in better agreement with the simulation results (c.f. Fig. \ref{Lalpha vs M}). 
Equation \ref{Egrav max} is thus really an upper bound that we do not expect to be saturated in this regime.\\ \\
In addition to gravitational potential energy, there is an additional source of power that can be converted into \Lya~photons even in the case of pure accretion, without feedback from stars or AGN: the cosmic ionizing background. 
An upper bound for how much power the ionizing background can contribute to \Lya~cooling within a dark matter halo can be estimated as the total inward flux of energy from ionizing photons across a sphere of a virial radius. 
A more realistic estimate is obtained by multiplying this quantity by a factor $f_{\rm cov}$ accounting for the fact that only a small fraction of these ionizing photons are actually absorbed within the virial shell:
\begin{equation}
\dot{E}_{\rm ion} = 4 \pi r_{\rm vir}^{2} F_{\rm ion} f_{\rm cov},
\end{equation}
where $F_{\rm ion}=\pi \int_{\nu_{\rm HI}}^{\infty} d\nu J_{\rm \nu}$ and $J_{\rm \nu}$ is the specific intensity of the ionizing background, assumed homogeneous and isotropic, which is a good assumption just above the hydrogen ionization edge at $\nu_{\rm HI}$ \citep[e.g.,][]{2009ApJ...703.1416F}. 
Since the background spectrum has a strong absorption edge at 4$\nu_{\rm HI}$ owing to intergalactic HeII absorption, we can optimistically take $J_{\nu} \approx J_{\nu_{\rm HI}}$ for $\nu \in [\nu_{\rm HI},~4\nu_{\rm HI}]$, and $J_{\nu}=0$ beyond $4\nu_{\rm HI}$. Then:
\begin{equation}
\dot{E}_{\rm ion} \approx 12 \pi^{2} r_{\rm vir}^{2} \nu_{\rm HI} J_{\nu_{\rm HI}} f_{\rm cov},
\end{equation}
For the hydrogen photoionization rate $\Gamma=0.6\times10^{-12}$ s$^{-1}$ measured from the \Lya~forest at $z=3$ \citep[][]{2008ApJ...688...85F}, this spectrum implies $J_{\nu_{\rm HI}}=1.5\times10^{-22}$ erg s$^{-1}$ cm$^{-2}$ Hz$^{-1}$. Expressing in terms of halo mass, this gives
\begin{equation}
\label{Edot ion}
\dot{E}_{\rm ion} \approx 1.9 \times 10^{41} {\rm~erg~s^{-1}} 
\left(
\frac{
M_{\rm h}
}
{10^{12}~{\rm M}_{\odot}}
\right)^{2/3}
\left(
\frac{4}{1+z}
\right)^{2}
\left(
\frac{J_{\nu_{\rm HI}}}{1.5\times10^{-22}~{\rm erg~s^{-1}~cm^{-2}~Hz^{-1}}}
\right)
\left(
\frac{f_{\rm cov}}{0.05}
\right).
\end{equation}\\
Comparison of equations \ref{Egrav max} and \ref{Edot ion} suggests that photoionization from the cosmic background can account for only about a percent of the maximum power available from the release of gravitational potential energy at the fiducial halo mass $M_{\rm h}=10^{12}$ M$_{\odot}$. 
Owing to the different mass dependences, and because our numerical results indicate that the gravitational power upper bound can be far from saturated in some plausible cases (\S \ref{self shielding}), we however cannot exclude that the ionizing background contributes a significant fraction of the \Lya~cooling emission in some regimes. 
This contribution is included in our simulations, though, and while it is difficult to disentangle from gravitational energy, circumstantial evidence suggests that it does not dominate. Indeed, in our most realistic simulations with a consistent self-shielding approximation the gravitational power upper bound is never systematically violated (see the right panel of Fig. \ref{Lalpha vs M}), and a simulation of our A1 halo with the ionizing background completely turned off yields a nearly equal cooling luminosity (within 30\%) at $z=3$  as our prescription 9 when excluding the multiphase star forming regions in the same manner.\\ \\
Although we intentionally ignore \Lya~photons produced by star formation throughout most of this work, it is interesting to analytically estimate the photoionization stellar \Lya~emission relative to pure cooling under simplified assumptions. 
Numerical simulations indicate that, in the absence of strong feedback, the star formation rate in high-redshift halos closely follows the cold gas accretion rate \citep[][]{2005MNRAS.363....2K, 2009MNRAS.395..160K, baryona}, i.e. $SFR\approx \dot{M}_{\rm cold}$, except for modest . 
Stellar \Lya~emission results from the conversion of the ionizing radiation from young stars via absorption and recombination by the surrounding medium, and therefore scales with the star formation rate, with a proportionality constant that depends on the initial mass function. 
For standard assumptions, $L_{\alpha}^{\rm SF} = 10^{42}$ erg s$^{-1}$ [SFR/(M$_{\odot}$ yr$^{-1}$)] \citep[e.g.,][]{1999ApJS..123....3L}. 
If we assume that $\dot{M}_{\rm cold} \approx \dot{M}_{\rm gas}$,
\footnote{This assumption is increasingly violated at high halo masses, where hot halo gas becomes more prevalent, but since the cooling \Lya~emission also scales with $\dot{M}_{\rm cold}$ instead of $\dot{M}_{\rm gas}$ to first order, the errors made should approximately cancel when taking the ratio of stellar to cooling \Lya~below.} then we can again make use of the average fitting formula in equation \ref{Mdot eq} to derive the average stellar Ly$\alpha$ luminosity as a function of halo mass and redshift:
\begin{equation}
\langle L_{\alpha}^{\rm SF} \rangle \approx 
2.1\times10^{44}~{\rm erg~s^{-1}} f_{\alpha,\rm esc}
\left( \frac{M}{10^{12}~{\rm M_{\odot}}} \right)^{1.4}
\left( \frac{1+z}{4} \right)^{2.5} 
\left( \frac{f_{\rm gas}}{0.165} \right),
\end{equation}
where the $f_{\alpha,\rm esc}$ factor quantifies the fraction of \Lya~photons that avoid destruction by dust. 
We can then combine this with equation \ref{Egrav max} to find the characteristic ratio of stellar to cooling \Lya:
\begin{equation}
\label{stellar to cooling Lya}
\frac{\langle L_{\alpha}^{\rm SF} \rangle}{\langle \dot{E}_{\rm grav }\rangle}
\approx
5.53 \left( \frac{f_{\alpha,\rm esc}}{f_{\rm eff}} \right)
\left( \frac{M}{10^{12}~{\rm M_{\odot}}} \right)^{-0.4}
\left( \frac{1+z}{4} \right)^{-1}
\frac{5.2}{c/[\ln{(1+c)}-c/(1+c)]}.
\end{equation}
The main point to take away is that it is a priori difficult to predict whether stellar \Lya~emission should dominate over cooling \Lya, since the ratio depends on a number of uncertain parameters, including $f_{\alpha,\rm esc}$ and $f_{\rm eff}$. 
Furthermore, the ratio depends on both halo mass and redshift, so that theoretical predictions based on model galaxies at fixed $M$ and $z$ \citep[e.g.,][]{2009ApJ...696..853L, 2009ApJ...704.1640L} cannot be universally applied. The actual ratio of stellar ionization to cooling \Lya~is likely to be somewhat suppressed relative to the simple scaling in equation \ref{stellar to cooling Lya} because galactic winds can remove a large fraction of the gas from the star-forming reservoir \citep[e.g.,][]{2005ApJ...618..569M, 2010MNRAS.406.2325O}, therefore suppressing the \Lya~emission from star formation ionization, without significantly affecting the accretion of cold material and hence the cooling \Lya.

\section{B. IONIZING RADIATIVE TRANSFER}
\label{ionizing radiative transfer}
The opacity in the \Lya~line (eq. \ref{lya tau0}) and the \Lya~emissivity (eq. \ref{recombinations eq} and \ref{collisions eq}) depend on the ionization state of the hydrogen atoms.
The SPH code GADGET calculates the ionization of gas elements assuming photoionization equilibrium with a uniform UV background \citep[e.g.,][]{1996ApJ...461...20H, 2009ApJ...703.1416F} (see \S \ref{ionization and thermal}).
This optically thin treatment misses the effects of self-shielding in dense regions (\S \ref{ionization and thermal}). 
To correct for this, we use a post-processing method to solve the radiative transfer problem on the gas distribution provided by GADGET to obtain more realistic ionization fractions and to identify the regions that self-shield from external radiation. 
The basic assumption is that the gas dynamics does not significantly depend on the radiative transfer effects missed in the SPH calculation. 
However, re-simulations are also used to approximate the dynamical and thermodynamical response to self-shielding (\S \ref{on the fly self shielding}). 
\\ \\
We use a ray tracing algorithm on the same grid as the \Lya~radiative transfer (Appendix \ref{rt code}).
Specifically, we send a normal ray from each cell on the surface of the radiative transfer volume along the inward direction ($x+$, $x-$, $y+$, $y-$, $z+$, or $z-$) and iteratively solve for the equilibrium ionization structure.
For this calculation, we take the gas temperatures provided by the SPH calculation and assume that all the helium is in the form of HeIII. 
These are good approximations for the gas outside of self-shielded regions and they therefore correctly capture the onset of self-shielding. 
In principle, the distribution of helium ionization affects the predicted \Lya~luminosity through the abundance of free electrons. 
The exact ionization state of helium can however matter significantly only inside self-shielded gas, since the free electron number density is dominated by hydrogen elsewhere. 
Assuming that all the helium is in HeIII overestimates the number of free electrons and hence the recombination and collisional excitation rates. 
However, the overestimation is only significant for temperatures $T\lesssim10^{4}$ K, at which the self-shielded gas is found to contribute negligibly to the total \Lya~emission (\S \ref{self shielding}). 
During the re-simulations, turning off the ionizing background in dense gas results in most of the helium being in the form of HeI in self-shielded gas, as should be the case since HeI traces HI for realistic ionizing spectra. 
Errors associated with the helium ionization state are therefore expected to be small in all cases. 
The rays are assumed to originate from a diffuse ionizing background with hydrogen photoionization rate $\Gamma_{\rm HI}^{\rm bkg}$.
For simplicity, we only update the hydrogen ionization fractions and only keep track of the photoionization rate $\Gamma_{{\rm HI},i}$ along each ray, rather than the full ionizing spectrum.
This is a fair approximation because of the small thickness of the self-shielding layer, in which the gas transitions from almost fully ionized to almost completely neutral.
Indeed, the thickness of this layer is approximately equal to the mean free path of ionizing photons within it:
\begin{equation}
\Delta l_{\rm ss} \sim \Delta l_{\rm mfp} \sim \frac{1}{n_{\rm HI} \sigma_{\rm HI}(\nu_{\rm HI})} \approx 5~{\rm pc} \left( \frac{n_{\rm HI}}{\rm 0.01~cm^{-3}} \right)^{-1},
\end{equation}
where $\sigma_{\rm HI}$ is the hydrogen photoionization cross section and $\nu_{\rm HI}$ is its ionization frequency, corresponding to 13.6 eV.
This mean free path is generally much smaller than the spatial resolution of both our SPH simulations (Table \ref{simulations table}) and our radiative transfer grids. 
One circumstance in which the self-shielding layer could be significantly thicker would be the case in which the illumination is dominated by a nearby quasar \citep[e.g.,][]{2008ApJ...672...48C, 2010ApJ...708.1048K}; we do not attempt to model this here, although it would be interesting to explore in the future.\\ \\
Along each ray $i$, the optical depth from the background is calculated assuming that all the ionizing photons effectively have a frequency $\nu_{\rm HI}$, $\tau_{i} = \int dl n_{\rm HI} \sigma_{\rm HI}(\nu_{\rm HI})$, and the photoionization rate is correspondingly attenuated as $\Gamma_{{\rm HI},i} = \Gamma_{\rm HI}^{\rm bkg} e^{-\tau_{i}}$.
The equilibrium solution is obtained iteratively as follows.
Trial values for the ionized fraction of hydrogen in each cell are first arbitrarily chosen.
The optical depth from the background to each cell, along each normal direction, is then computed.
The effective photoionization rate within the cell is taken to be the average of the attenuated value from each direction, $\Gamma_{\rm HI} \equiv (1/6) \sum_{i} \Gamma_{{\rm HI},i}$.
Using this photoionization rate, an updated hydrogen fraction is calculated using the balance equation \ref{ionization balance}.
For this step, the electron number density $n_{e}$ is taken equal to its value after the previous update.
The procedure is repeated until the ionization fraction has converged in each cell.
The convergence criterion is set to one part in $100$.\\ \\
In reality, the gas temperatures provided by the basic hydrodynamical simulations are in error within self-shielded regions. 
We thus store the mean attenuation from the background, $\langle e^{-\tau_{i}} \rangle$, in each cell as an indicator of self-shielding.
Different prescriptions for the self-shielded gas and re-simulation procedures are explored in \S \ref{self shielding}.

\section{C. Ly$\alpha$ RADIATIVE TRANSFER: $\alpha$RT}
\label{rt code}
In this Appendix, we describe the new three-dimensional \Lya~radiative transfer code, called ``$\alpha RT$'', developed for this project.

\subsection{Basic Physics and Algorithm}
\label{basic physics and algorithm}
The basic algorithm is the standard Monte Carlo method, which follows the scatters of a prescribed number, $N_{\rm ph}$, of \Lya~photons and statistically estimates the emergent morphology and line profile from them \citep[e.g.,][]{2002ApJ...578...33Z, 2005ApJ...628...61C, 2006ApJ...649...14D, 2006A&A...460..397V, 2006ApJ...645..792T, 2007ApJ...657L..69L, 2009ApJ...696..853L}.
The principal improvements over many of these codes are that it is fully three-dimensional and its implementation is parallel to take advantage of distributed computing resources.
The code therefore scales well to large problems with complex geometries and so can be run on the outputs of large-scale hydrodynamical simulations.
It produces \Lya~images, spectra, as well as integral field images (spectra as a function of position on the sky).
As it was independently developed, its results provide useful checks of existing numerical calculations. 
It has been demonstrated \citep[][]{2006ApJ...645..792T, 2009ApJ...696..853L} that the Monte Carlo algorithm can be combined with mesh refinement techniques relatively straightforwardly, which is particularly valuable for problems of large dynamic range, for instance when attempting to capture the details of scattering within the clumpy ISM of galaxies at the same time as the large scale cosmological radiative transport. 
We plan to build on the infrastructure developed in this work and implement adaptive refinement in future versions of the code. 
Here, we briefly outline the basic physics and methodology of the current implementation.
Our notation follows that of \cite{2006ApJ...649...14D}.\\ \\
There are three relevant reference frames: the frame of the observer, the fluid frame, and the frame of a particular atom, which may have thermal motion with respect to the fluid frame.
The observer is assumed to be located at $z=+\infty$ and to be stationary with respect to the simulation volume, apart from a possible Hubble flow that redshifts all the photons from the simulation volume by the same factor.
We assume that the simulation volume is sufficiently small that its volume is essentially at a single redshift.
A quantity $Q$ in the observer's frame is denoted $Q'$ in the fluid frame and $\tilde{Q}$ in the frame of the atom.\\ \\
The Monte Carlo photons are injected with a probability per unit volume proportional to the local \Lya~emissivity (\S \ref{lya emission}).
The photons are assumed to have energy exactly at the line center in the frame moving with the fluid and the emission is taken to be isotropic.
The natural and thermal broadening of the emission line have a negligible impact on the results, as the resonant scatters rapidly erase their memory.
The \Lya~photons are then each transported by simulating their resonant scatters.\\ \\
The \Lya~optical depth through a HI column density $N_{\rm HI}$ for a photon at the line center is given by
\begin{equation}
\label{lya tau0}
\tau_{0} = N_{\rm HI} \sigma_{\Lya}(\nu_{0}) =
N_{\rm HI} f_{12} 
\frac{ \sqrt{\pi} e^2 }
{m_{e} c \Delta \nu_{\rm D}}
\approx
8.3\times10^{6}
\left(
\frac{N_{\rm HI}}{2\times10^{20}~{\rm cm^{-2}}}
\right)
\left(
\frac{T}{2\times10^{4}~{\rm K}}
\right)^{-1/2},
\end{equation}
where $\sigma_{\Lya}$ is the \Lya~cross section, $\nu_{0}$ is the \Lya~line center, $f_{12}$ is the oscillator strength of the transition, $e$ is the charge of the electron, $m_{e}$ is its mass, $c$ is the speed of light, and $\Delta \nu_{\rm D} = v_{\rm th} \nu_{0}/c$, with $v_{\rm th}=(2 k_{\rm B} T/m_{\rm p})^{1/2}$.
It is convenient to locally define a dimensionless frequency $x \equiv (\nu-\nu_{0})/\Delta \nu_{\rm D}$.
The optical depth for photons of arbitrary energy is then
\begin{equation}
\tau_{x} = \tau_{0} H(a,~x) = \tau_{0}
\frac{a}{\pi}
\int_{-\infty}^{\infty}
\frac{e^{-y^2}dy}{(y-x)^2+a^2},
\end{equation}
where $a=A_{21}/(4\pi \Delta \nu_{\rm D})$ is the Voigt parameter and $A_{21}$ is the \Lya~Einstein A coefficient.\\ \\
In the absence of perturbations, the absorption of a \Lya~photon is quickly followed (after $A_{21}^{-1}\sim10^{-9}$ s) by the reemission of another \Lya~photon of the same energy.
Owing to the motion of the atom, the scatter is however not coherent in the observer's frame.
Defining ${\bf v}_{\rm a}\equiv {\bf v}_{\rm bulk} + {\bf v}_{\rm th}$ to be the net velocity of the scattering atom, the frequency of a photon of incident frequency $x_{i}$ after scattering is
\begin{equation}
\label{scattered frequency}
x_{o} = x_{i} + \frac{{\bf v}_{\rm a} \cdot ({\bf k}_{i}-{\bf k}_{o})}{v_{\rm th}} + g ({\bf k}_{i} \cdot {\bf k}_{o} - 1) 
+ {\mathcal O}
\left( 
\left(
\frac{v_{\rm th}}
{c}
\right)^{2} 
\right),
\end{equation}
where ${\bf k}_{i}$ and ${\bf k}_{o}$ are unit vectors in the directions of the incident and outgoing photons, respectively \citep[e.g.,][]{2006ApJ...649...14D}. 
The Hubble expansion is modeled by including a component ${\bf v_{\rm H}} = H(z) {\bf r}$, where ${\bf r}$ is the displacement vector from the center of the system and $H(z)$ is the Hubble parameter, in the bulk velocity term. 
The $g$ ``recoil'' term accounts for the average transfer of momentum from the photon to the atom during the scatter and can be written as $g=h \Delta \nu_{\rm D}/2k_{\rm B}T$ \citep[][]{1959ApJ...129..551F}.
The effect of recoil is usually small \citep[][]{1971ApJ...168..575A} and is ignored in our calculations.\\ \\
Because the optical depth near the \Lya~line center is very large in astrophysical conditions, a \Lya~photon typically travels only a short distance before being scattered.
The numerous scatters result in a random walk in space.
At the same time, the scatters cause the photon to oscillate in frequency and it usually escapes the medium during an excursion far into a damping wing of the line, where the optical depth is greatly reduced \citep[e.g.,][]{1949BAN....11....1Z, 1952PASJ....4..100U, 1959ApJ...129..551F, 1962ApJ...135..195O, 1972ApJ...174..439A, 1973MNRAS.162...43H, 1990ApJ...350..216N, 1996ApJ...468..462G}.
As a result, the emergent \Lya~line profile is heavily affected by the medium through which it propagates and therefore provides a probe of the latter.\\ \\
To simulate the scatters of the Monte Carlo photons, we at each step randomly pick the optical depth $\tau$ before the next scatter.
By definition, $\tau$ has a PDF $P(\tau) \propto e^{-\tau}$.
Given the frequency and propagation direction of the photon, we integrate through the medium until an optical depth $\tau$ has been reached; this defines the position of the next scatter.
The frequency seen by a fluid element along the way in general differs from that seen by the observer and is given by the Doppler shift formula $x_{i}'=x_{i} - {\bf k_{i}}\cdot{\bf v}_{\rm bulk}/v_{\rm th}$.
The frequency of the outgoing photon is then calculated using equation \ref{scattered frequency}, which requires picking ${\bf k}_{o}$ and ${\bf v}_{\rm th}$.\\ \\
The direction of the outgoing photon is picked from a dipolar phase function, $P({\bf k}_{i} \cdot {\bf k}_{0}) \propto 1 + ({\bf k}_{i} \cdot {\bf k}_{0})^{2}$.
Owing to quantum mechanical effects, this Rayleigh phase function is strictly valid only for scatters with $\tilde{x}_{i}\gtrsim0.2$ \citep[see][and references therein]{2008MNRAS.386..492D}, but the multiple scatters act to quickly randomize propagation directions so that the precise phase function is unimportant.
An isotropic phase function, $P({\bf k}_{i} \cdot {\bf k}_{0}) \propto {\rm const}$, gives identical results in practically all astrophysical conditions, aside for polarization, which we neglect here.\\ \\
The first part of the scattering atom velocity is simply the local bulk velocity of the fluid, ${\bf v}_{\rm bulk}$.
To this, a thermal velocity ${\bf v}_{\rm th}$ must be added.
In a frame moving with the fluid, but oriented such that the $z'$ axis is parallel to ${\bf k}_{i}$, the thermal velocity can be expressed as $v_{\rm th}(u_{\perp1}',~u_{\perp2}',~u_{||}'$), where $u_{||}'$ is the normalized component parallel to the direction of propagation, and $u_{\perp1,2}'$ are perpendicular components.
The probability of scattering off an atom with parallel component $u_{||}'$ is the product of the one-dimensional thermal distribution and the absorption cross section,
\begin{equation}
\label{P u par}
P(u_{||}') = \frac{a}{\pi}
\frac{e^{-u_{||}'^{2}}}
{(x_{i}'-u_{||}')^{2}+a^{2}}
H^{-1}(a,~x_{i}'),
\end{equation}
while $u_{\perp1,2}'$ are simply thermally distributed since the probability of absorption is independent of these normal components.\\ \\
The procedure is repeated for each scatter of each Monte Carlo until it escapes the computational volume.
In this work, we assume that all photons escape the medium but the algorithm can be straightforwardly extended to model photon destruction (e.g., by dust) by specifying a destruction probability at each integration step or scattering event.

\subsection{Numerical Aspects}
\label{numerical aspects}
Having described the essential aspects of the algorithm, we now elaborate on numerical aspects of its implementation relevant to its accuracy and performance.\\ \\
The generation of random numbers is a key element of the Monte Carlo method.
In most instances, these are straightforwardly generated using either a transformation method or a rejection method \citep[e.g.,][]{1992nrca.book.....P}.
An important exception is the generation of $u_{||}'$, whose PDF (eq. \ref{P u par}) is not analytically integrable and for which there is not an obvious efficient bounding function to use in the rejection method.
Since a variate from this distribution is generated at each scatter, it is important to have a fast implementation.
We generalize the rejection method described in the appendix of \cite{2002ApJ...568L..71Z} to use three critical velocities (in their notation, $u_{1}$ and $u_{2}$ in addition to $u_{0}$) in order to minimize the number of rejections.\\ \\
Although the distance between neighboring grid points is $\Delta l \equiv L/N_{\rm p}$, photon trajectories are traced exactly in the sense that the photon can be located at any coordinate $(x,~y,~z)$ in the box along its ray.
This can greatly improve the accuracy of the calculations possible with limited computational resources.
For instance, a large and/or very optically thick medium may be well described by a relatively coarse grid if it is fairly homogeneous.
However, because of the short mean free path of the \Lya~photons, the radiative transfer would be poorly captured if the position of the photon were restricted to widely separated grid points.
The integration along rays is still done in finite steps of length $\Delta l_{\rm int}$, but this parameter can be specified arbitrarily and in particular can be $\ll \Delta l$.
Since the photon position is in general between the grid points on which physical properties are stored, these must be interpolated onto the ray.
Our code can use either nearest grid point (NGP) or cloud-in-cell (CIC) interpolation \citep[e.g.,][]{1988csup.book.....H}.
While CIC is more accurate, NGP produces similar results but requires a factor of several fewer operations per integration step. 
We use NGP interpolation in the calculations presented in this work.\\ \\
Finally, we use an ``accelerated scheme'' \citep[e.g.,][]{2002ApJ...567..922A} that greatly speeds up radiative transfer calculations by skipping scatters in the core of the \Lya~line, during which the photon is nearly stationary in space.
Specifically, we define a critical frequency $x_{\rm crit}$ and say that a photon with $|x'|<x_{\rm crit}$ is in the core of the line.
Photons in the line core are then forced into the wings by allowing them to scatter only off high-velocity atoms.
This is achieved by truncating the thermal Gaussian distribution from which $u_{\perp1}$ and $u_{\perp2}$ are drawn to zero for $|u|<x_{\rm crit}$.
Skipping core scatters effectively sets the mean free path of core photons to zero.
In the limit $x_{\rm crit} \to 0$, the algorithm becomes exact and we can vary $x_{\rm crit}$ to verify the convergence of our calculations.
In practice, we find that $x_{\rm crit}=3$ generally speeds up the calculations (by as much as orders of magnitude in the very optically thick regime) while producing faithful results.
We use this value as default in this work. 
Following \cite{2006ApJ...648..762T, 2006ApJ...645..792T} and \cite{2009ApJ...696..853L}, we further speed up the code by making use of the fact that in extremely optically thick radiative transfer cells (which, in the discretized numerical scheme, are internally static and uniform), the solution to the radiative transfer problem is well approximated by a rescaled version of the analytic solution obtained by \cite{1990ApJ...350..216N}. 
Specifically, the explicit scatters in cells with $a \tau_{0} \gtrsim 10^{3}$ can be skipped by choosing the emerging frequency according the known analytic solution.

\begin{figure*}[ht] 
\centering
\mbox{
{\includegraphics[width=0.5\textwidth]{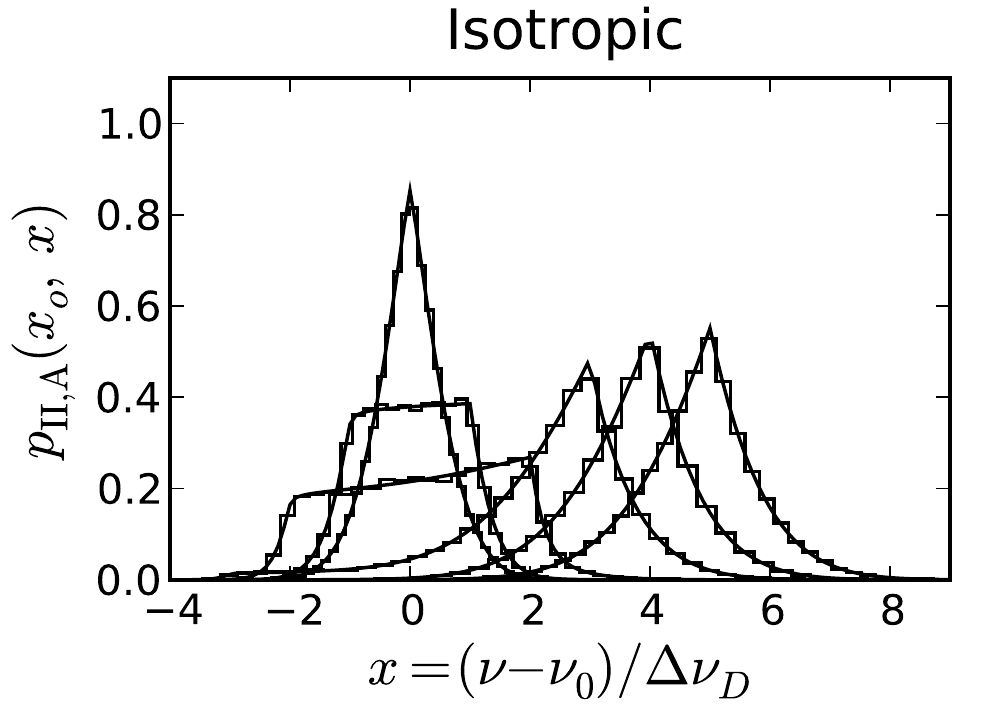}}
{\includegraphics[width=0.5\textwidth]{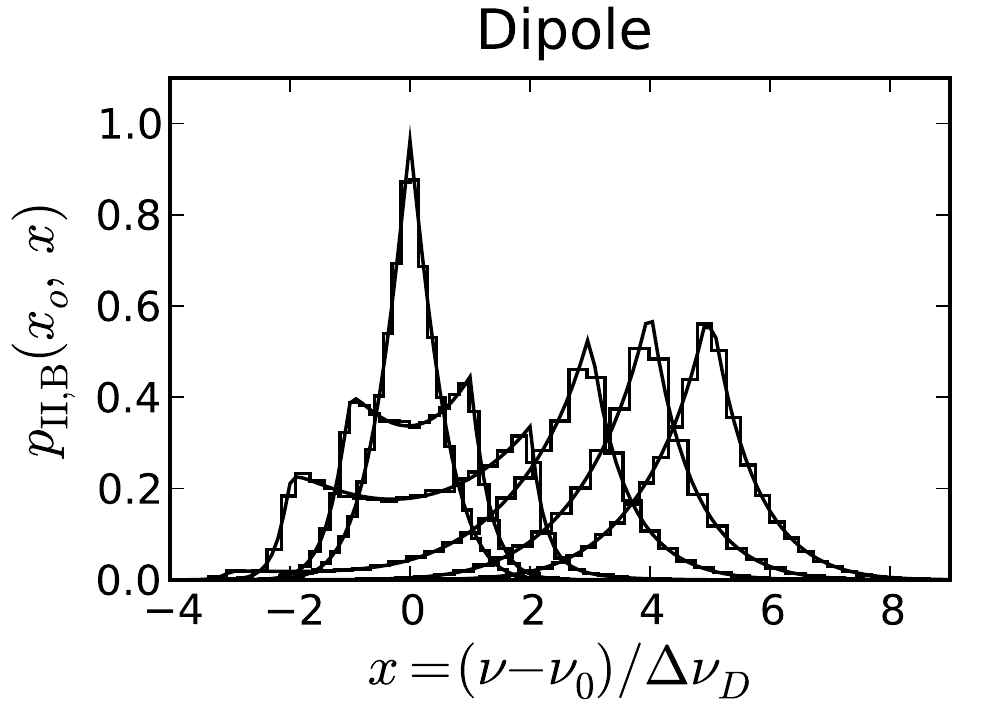}}
}\\
\caption{Redistribution function test: PDF of the outgoing dimensionless frequency $x$ as a function of the incident $x_{0}$, for $x_{0}=0,~1,~2,~3,~4,{\rm~and~}~5$ from left to right in each panel.
The histograms show the PDFs estimated by our Monte Carlo code and the smooth curves show the corresponding analytic solutions.
The left panel shows the case of an isotropic redistribution function and the right panel shows the case of a dipole redistribution function.}
\label{redistribution test figure} 
\end{figure*}

\subsection{Visualization}
\label{visualization}
The principal application of our code is to produce images and spectra that can be compared to observations.
To do so, the code keeps track of a three-dimensional ``integral field'' array, of dimensions $N_{\rm g}^{2} \times N_{\nu}$.
The first two dimensions correspond to the projected position on the sky as viewed by the observer, or pixels.
The third dimension divides each pixel into $N_{\nu}$ frequency bins between specified boundaries.
From the full integral field array, with a spectrum at each pixel, a \Lya~image can be then be constructed by summing over frequency bins within each pixel.
A spectrum toward any particular direction, or along an arbitrarily placed slit, can be produced by considering only the relevant pixels.\\ \\
To construct the integral field array, a number proportional to the probability that the outgoing photon escapes directly toward the observer is added to the corresponding pixel and frequency bin at each scatter.
Since the observer is assumed to be located at $z=\infty$, we take this numerical increment to be
\begin{equation}
\frac{3}{16\pi} [1 + ({\bf k}_{i} \cdot {\bf z})^{2}] e^{-\tau_{z}}.
\end{equation}
The factor in the square brackets accounts for the probability that the photon is reemitted toward the observer and $e^{-\tau_{z}}$ is the probability of escape in a single flight, where $\tau_{z}$ is the optical depth to $z=\infty$, given the frequency that the photon would have had if it had been scattered in that direction.
This procedure \citep[see also][]{2002ApJ...578...33Z, 2005ApJ...628...61C} is necessary because only an infinitesimal fraction of the Monte Carlo photons actually emerge exactly in the $z=\infty$ direction and is termed the ``next-event estimator'' \citep[][]{dupree}.
To convert the dimensionless counter to observed physical intensity, we multiply by $L_{\alpha} /[ N_{\rm ph} (\Delta l)^{2}(1+z)^{4}]$, where $L_{\alpha}$ is the total \Lya~luminosity and the $z$ factor accounts for cosmological surface brightness dimming. 
We have verified that this next-event estimator conserves the number of photons in isotropic cases, i.e. that the luminosity inferred from the resultant image is equal to the luminosity of the central source, from which only a small fraction of the emitted photons directly reach the observer. 

\begin{figure*}[ht] 
\centering
\mbox{
{\includegraphics[width=0.5\textwidth]{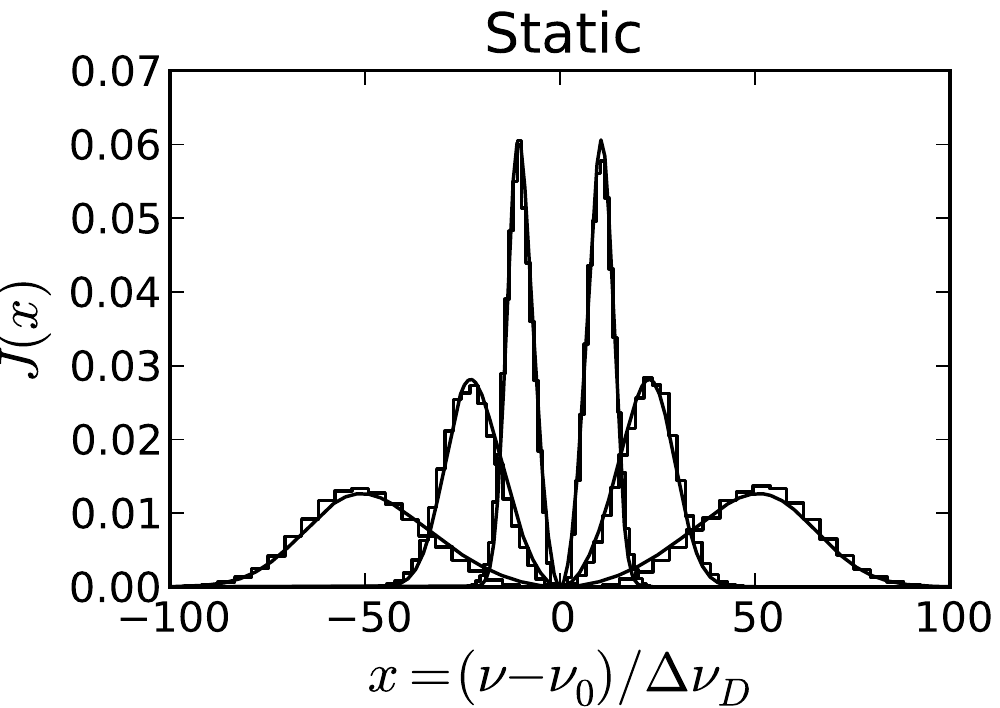}}
}\\
\caption{Static sphere test: Emergent line profile for a monochromatic source at the center of a uniform static sphere for optical depths at the line center $\tau_{0}=10^{5},~10^{6},~{\rm~and~}10^{8}$.
The histograms show the numerical solutions obtained with our code and the smooth curves show the corresponding analytical solutions following \cite{2006ApJ...649...14D} (eq. \ref{dijkstra solution}).
The numerical solution was computed for a temperature $T=10$ K.}
\label{static test figure} 
\end{figure*}

\begin{figure*}[ht] 
\centering
\mbox{
{\includegraphics[width=0.5\textwidth]{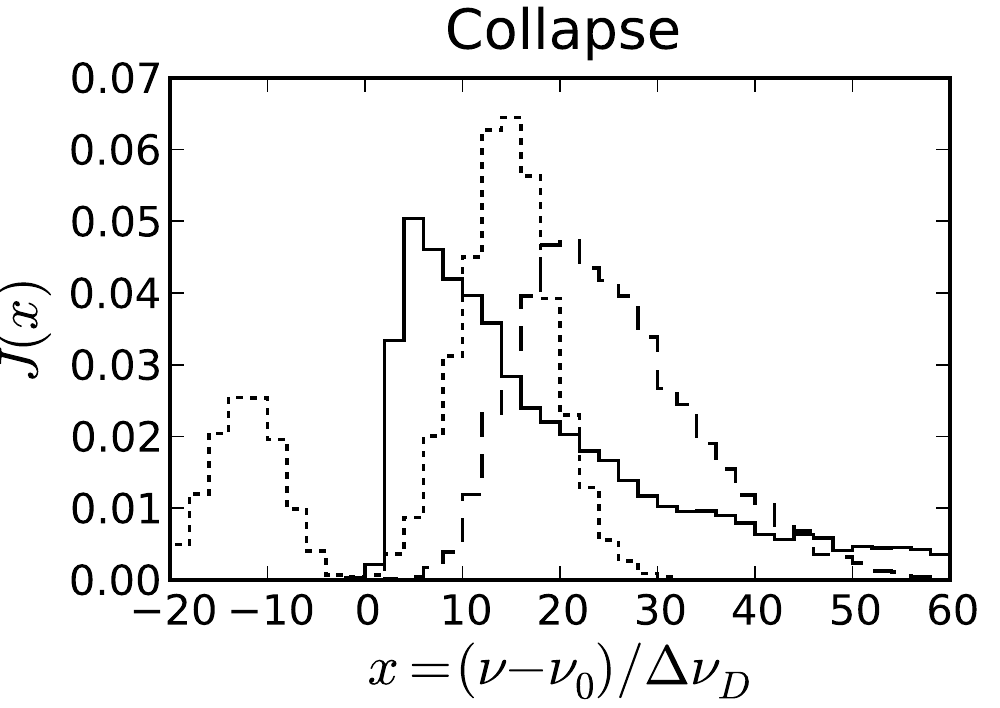}}
{\includegraphics[width=0.5\textwidth]{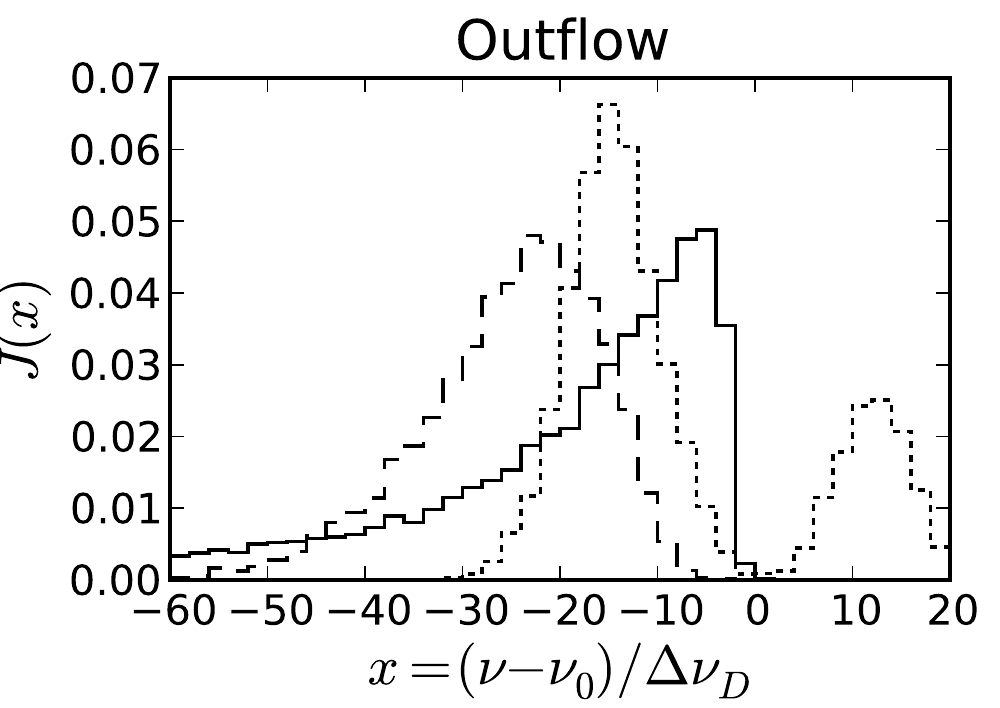}}
}\\
\caption{Collapsing and outflowing sphere tests.
The left panel shows the numerical results for a collapsing uniform sphere of radius $R$, neutral hydrogen column density $N_{\rm HI}=2\times10^{20}$ cm$^{-2}$ from the center to the surface, and temperature $T=20,000$ K.
In each case, the sphere is collapsing with a velocity ${\bf v}=(-r/R) v_{\rm max} {\bf \hat{r}}$, with $v_{\rm max}=20$ (dotted), 200 (dashed), and 2,000 km s$^{-1}$ (solid).
The right panel shows the same results for an expanding sphere with the same parameters, but reversed velocity field.
The results can be compared with Figure 7 of \cite{2006A&A...460..397V} for the expanding case.
The collapsing case is symmetric under the replacement $x\to-x$.}
\label{dynamic test figure} 
\end{figure*}

\subsection{Test Problems}
\label{test problems}
In order to be confident in the results obtained from our radiative transfer code, it is important to test it on problems with known solutions. 
Figures \ref{redistribution test figure}, \ref{static test figure}, and \ref{dynamic test figure} show the results of tests of the $\alpha RT$ code for the following problems:
\begin{itemize}
\item {\bf Redistribution function.} The redistribution function, $R(x_{o}',~x_{i}')$, is the PDF of the outgoing photon dimensionless frequency, $x_{o}'$, as a function of the incoming photon frequency, $x_{i}'$, in the fluid frame.
Although not a solution to the radiative transfer problem \emph{per se}, the Monte Carlo algorithm effectively generates random variates from this distribution at each scatter when it picks ${\bf v}_{\rm th}$ and ${\bf k}_{o}$, and applies equation \ref{scattered frequency}.
A useful test of this key portion of the algorithm is thus to compare the Monte Carlo-generated redistribution function with analytically-derived expressions \citep[][]{1952PASJ....4..100U, 1962MNRAS.125...21H, 1974ApJ...192..465L}.
Figure \ref{redistribution test figure} compares the results of our code with the analytic solutions presented by \cite{1974ApJ...192..465L} (for a naturally-broadened intrinsic line) for the cases of isotropic and dipole phase functions.
To reproduce the redistribution function for core scatters (small $x_{i}'$), the accelerated scheme must be turned off ($x_{\rm crit}=0$).

\item {\bf Static sphere.}
An analytic solution to the \Lya~radiative transfer problem exists for the case of a monochromatic point source at the center of an extremely optically thick, uniform, and static sphere of line-center optical depth $\tau_{0}$ from the source to the surface:
\begin{equation}
\label{dijkstra solution}
J(x) = \frac{\pi^{3/2}}{\sqrt{6}a \tau_{0}}
\left\{
\frac{x^{2}}{1 + \cosh{[\sqrt{2\pi^{3}/27}(|x|^{3}/a\tau_{0})]}}
\right\}
\end{equation}
\citep[][]{2006ApJ...649...14D}.
This solution is a generalization of the plane-parallel results obtained by \cite{1973MNRAS.162...43H} and \cite{1990ApJ...350..216N}.
It differs by a factor of $2\pi$ from the expression in \cite{2006ApJ...649...14D} because it is normalized to integrate to unity here.
Figure \ref{static test figure} compares the results obtained with our code for $\tau_{0}=10^{5},~10^{6},{\rm~and~}10^{7}$ and a gas temperature $T=10$ K with this solution.

\item {\bf Collapsing/outflowing sphere.}
Velocity fields are crucial to the \Lya~radiative transfer physics, as Doppler shifts modify the perceived optical depth and can open ``paths of least resistance'' through which photons can escape the medium.
It is therefore important to test the code in dynamic situations.
Unfortunately, few analytic results with velocity fields are available.
We therefore instead compare our numerical solutions against those obtained with other codes.
Specifically, we consider the following cases computed by \cite{2006A&A...460..397V} \citep[see also][]{2002ApJ...578...33Z, 2006ApJ...649...14D}:\\ \\
- A collapsing uniform sphere of radius $R$ with monochromatic central source: $N_{\rm  HI}=2\times10^{20}$ cm$^{-2}$ from the center to the surface, $T=20,000$ K, ${\bf v}=-(r/R)v_{\rm max} {\bf \hat{r}}$ with $v_{\rm max}=20,~200,~{\rm and~}2,000$ km s$^{-1}$.\\ \\
- Same, but expanding with ${\bf v}=(r/R)v_{\rm max} \hat{{\bf r}}$ for $v_{\rm max}=20,~200,~{\rm and~}2,000$ km s$^{-1}$.
\end{itemize}
The numerical solutions obtained with our code shown in Figure \ref{dynamic test figure} agree well with Figure 7 of \cite{2006A&A...460..397V} for the expanding case.
The collapsing case is symmetric under the replacement $x\to-x$.
 
\bibliography{references} 
 
\end{document}